\documentclass[reqno,12pt]{amsart}
\usepackage{fullpage}
\usepackage{amsfonts}
\usepackage{amssymb}
\usepackage{latexsym}
\usepackage{mathrsfs}
\usepackage[all]{xy}
\usepackage{multicol}
\usepackage{delarray}
\numberwithin{equation}{section}
\newcommand{\p}{\partial}

\def\eos/{equation of state}
\def\esos/{equations of state}
\def\tr{{\rm tr}}

\begin{document}

\title{Classification of conservation laws of compressible isentropic fluid flow in $n>1$ spatial dimensions}

\author{
Stephen C. Anco${}^1$ 
\lowercase{\scshape{and}} 
Amanullah Dar${}^{1,2}$ \\
\\\lowercase{\scshape{
${}^1$Department of Mathematics, Brock University, 
St. Catharines, ON Canada}}\\
\\\lowercase{\scshape{
${}^2$Department of Mathematics, Quaid-e-Azam University, 
Islamabad, Pakistan }}
}


\email{sanco@brocku.ca}
\email{amanullahdar@hotmail.com}

\thanks{S.C.A. is supported by an NSERC research grant. A.D. thanks HEC, Pakistan, for providing a 6-month fellowship grant and the Department of Mathematics at Brock University for additional support during the extended period of a research visit when this paper was completed.}

\begin{abstract}
For the Euler equations governing compressible isentropic fluid flow 
with a barotropic equation of state (where pressure is a function only of the density), 
local conservation laws in $n>1$ spatial dimensions are fully classified in 
two primary cases of physical and analytical interest: 
(1) kinematic conserved densities that depend only on the fluid density and velocity, in addition to the time and space coordinates; 
(2) vorticity conserved densities that have an essential dependence on the curl of the fluid velocity. 
A main result of the classification in the kinematic case is that 
the only equation of state found to be distinguished by admitting extra $n$-dimensional conserved integrals, 
apart from mass, momentum, energy, angular momentum and Galilean momentum (which are admitted for all equations of state), 
is the well-known polytropic equation of state with dimension-dependent exponent $\gamma=1+2/n$. 
In the vorticity case, no distinguished equations of state are found to arise, 
and here the main result of the classification is that, 
in all even dimensions $n\geq 2$, 
a generalized version of Kelvin's two-dimensional circulation theorem is obtained for a general equation of state. 
\end{abstract}
\maketitle

\section{Introduction}

Conservation laws and Hamiltonian structures 
are central to the mathematical study of fluid flow 
and have long been known for both the incompressible (ideal fluid) 
and compressible (inviscid fluid) Euler equations governing fluid flow
in two and three dimensions. 
Over the past few decades there has been considerable mathematical interest 
in studying the Eulerian fluid equations in $n$ dimensions \cite{AK}.

One strong motivation came from the work of Arnold \cite{Arn1966,Arn1969}
showing that the Euler equations for incompressible fluids
in $n$-dimensional spatial domains
have an elegant geometric formulation 
as the geodesic equation on the Lie group of volume-preserving diffeomorphisms
of the given domain of the fluid flow. 
This formulation gives an interesting geometrical significance to 
fluid conservation laws by interpreting them as geodesic first integrals 
related to invariance properties of the geodesic Lagrangian.
Subsequently, 
the main group-theoretic aspects of Arnold's work were extended to 
the compressible Euler equations, 
first \cite{GS,MRW} for isentropic fluids 
(whose entropy is constant throughout the fluid domain)
in which the pressure is specified to be a function only of density
as given by an \eos/,
then later \cite{Nov,AK} for adiabatic non-isentropic fluids 
(in which the entropy is conserved only along streamlines) 
where the pressure is given by a dynamical equation. 

The aim of the present paper and a sequel will be to give a complete picture of
the conservation laws of kinematic type and vorticity type
for general compressible fluids in $n>1$ dimensions for both isentropic and non-isentropic cases. 
By a {\it kinematic} conservation law we will mean 
a local continuity equation where the conserved density and flux 
depend only on the fluid velocity, pressure and density 
(but not their spatial derivatives), 
in addition to the time and space coordinates. 
Such conservation laws encompass 
the familiar physical continuity equations in two and three dimensions for  
mass, momentum and energy \cite{landau,Batch,chorin}. 
In contrast, a {\it vorticity} conservation law will refer to 
a local continuity equation for a conserved density and flux that have 
an essential dependence on the curl of the fluid velocity
in a form exhibiting odd parity under spatial reflections.
Examples of conservation laws with this form are 
helicity in three dimensions as well as 
circulation and enstrophy in two dimensions, which are well-known for incompressible fluid flow \cite{MB, She}. 

To-date all of the known $n$-dimensional fluid flow conservation laws \cite{Ibr1973,Ibr,KC} belong to these two classes but have been derived through special methods that fall short of providing a complete classification. An interesting open question we will settle in this paper for isentropic compressible fluid flow is to find all particular \esos/
for which the $n$-dimensional Eulerian fluid equations admit 
vorticity conservation laws or extra kinematic conservation laws.

In section 2, as preliminaries, the general formulation of local continuity 
equations and integral conservation laws for 
the Euler equations for compressible isentropic fluids 
in $n>1$ dimensions is reviewed. In particular, we introduce necessary and sufficient determining equations for directly finding conserved densities of any specified form. By solving the determining equations  
for kinematic conserved densities, we obtain a complete classification  
showing that apart from mass, momentum, and energy, the only additional conservation laws of kinematic form consist of Galilean momentum (connected with center-of-mass motion) and angular momentum holding for any \eos/, plus
dilational-type energies arising for polytropic \esos/
where the pressure is proportional to a particular dimension-dependent power of the density. 

Next in section 3 we solve the determining equations to find all vorticity conservation laws,
starting from the transport equation for the curl of the fluid velocity. 
This classification yields an odd-dimensional generalization of helicity
and an even-dimensional generalization of circulation and enstrophy,
which are found to hold for any \eos/. 
We show that the generalized circulation has an equivalent formulation as a constant of the fluid motion
defined on the boundary of any spatial domain that is transported in the fluid.
This new result gives a generalization of Kelvin's two-dimensional circulation theorem to all even dimensions $n\geq2$.

Finally, in section 4, we use the well-known Hamiltonian structure \cite{Ver} of 
the compressible Euler equations
to classify all Hamiltonian symmetries corresponding to 
the kinematic and vorticity conservation laws. 
In section 5 we make some concluding remarks, including a summary of our classification results stated in index notation.

A corresponding treatment of kinematic and vorticity conservation laws for adiabatic non-isentropic compressible fluids in $n>1$ spatial dimensions will be given in a separate paper \cite{paperII}.

\section{Compressible Isentropic Flows}

The Euler equations for compressible isentropic fluids in $\mathbb R^n$ (in the absence of external forces) with velocity $\mathbf{u}(t,\mathbf{x})$ and density $\rho(t,\mathbf{x})$ consist of 
\begin{eqnarray}
&&\p_t\mathbf{u}+\mathbf{u}\cdot\nabla\mathbf{u}+\frac{1}{\rho}\nabla p=0,
\label{veleqn}\\
&&\p_t\rho+\nabla\cdot(\rho\mathbf{u})=0,
\label{deneqn}
\end{eqnarray}
with a barotropic \eos/ for pressure
\begin{equation}
p=P(\rho).
\label{eos}
\end{equation}
Throughout, we will use bold notation to denote vector or tensor variables and operators. 
A dot will denote the Euclidean inner product as well as stand for contraction between vectors and tensors, 
while a wedge will denote the antisymmetric outer product of vectors and/or antisymmetric tensors. 

Fluid conservation laws are described by a local continuity equation \cite{Olv,AB2002b,BCA}
\begin{equation}
D_t T+D_{\mathbf{x}}\cdot\mathbf{X}=0
\label{conlaw}
\end{equation}
holding for all formal solutions of (\ref {veleqn})--(\ref{eos}), where $T$ and $\mathbf{X}$ are some functions of $t,\mathbf{x},\rho,\mathbf{u}$ and their $\mathbf x$-derivatives. 
Here $D_t$ and $D_\mathbf{x}$ denote total time and space derivatives respectively. 
Physically, $T$ is a conserved density with $\mathbf{X}$ being a corresponding spatial flux. In integral form, the continuity equation (\ref {conlaw}) is equivalently described by
\begin{equation}
\frac{d}{dt}\int_V T d^nx = -\int_{\p V}\mathbf{X}\cdot\hat{\mathbf{n}} d^{n-1}\sigma
\label{integral}
\end{equation}
where $V$ is any spatial domain in $\mathbb R^n$ through which the fluid is flowing and $\hat{\mathbf{n}}$ is the outward unit normal on the domain boundary $\p V$.
A physically more useful form 
for expressing fluid conservation laws (\ref {conlaw}) and (\ref{integral}) is obtained by considering a spatial domain $V(t)$ that moves with the fluid \cite{Ibr}. 
Then the flux through the moving boundary $\p V(t)$ is
${\boldsymbol\xi}=\mathbf{X}-T\mathbf{u}$ which is related to the conserved density $T$ by the transport equation
\begin{equation} 
D_t T+\mathbf{u}\cdot D_\mathbf{x}T=-(\nabla\cdot\mathbf{u})T-D_\mathbf{x}\cdot{\boldsymbol\xi}
\label{transeqn}
\end{equation}
where $D_t +\mathbf{u}\cdot D_\mathbf{x}$ represents the total convective (material) derivative and $\nabla\cdot\mathbf{u}$ represents the expansion or contraction of an infinitesimal volume moving with the fluid. 
The corresponding integral form of a fluid conservation law in a moving domain is accordingly expressed as
\begin{equation}
\frac{d}{dt}\int_{V(t)} T d^nx = -\int_{\p V(t)}{\boldsymbol\xi}\cdot\hat{\mathbf{n}} d^{n-1}\sigma
\label{moveqn}
\end{equation}
whereby $\int_{V(t)} T d^nx$ will be a constant of the fluid motion in $V(t)\subset\mathbb R^n$ if the net flux across $\p V(t)$ vanishes.

The determining equations for finding conserved densities $T$ are given by 
\begin{equation}
E_{\rho}({\mathcal D}_t T)=0=E_\mathbf{u}({\mathcal D}_t T)
\label{deteqn}
\end{equation}
where $E_{\rho}$ and $E_{\mathbf{u}}$ are spatial Euler operators \cite{AB2002b} with respect to $\rho$ and $\mathbf{u}$, and ${\mathcal D}_t$ is the total time derivative evaluated on solutions of the Euler equations (\ref{veleqn})--(\ref{eos}). 
(The explicit form of these operators is shown using index notation in section \ref{proof2}.) These equations (\ref{deteqn}) arise from the fact that spatial divergences $D_\mathbf{x}\cdot\mathbf{X} = -{\mathcal D}_t T$ have a characterization \cite{Olv,BCA} as functions of $t,\mathbf{x},\rho,\mathbf{u}$ and $\mathbf x$-derivatives of $\rho$,$\mathbf{u}$ that are annihilated by both of the spatial Euler operators.

A conservation law is locally trivial if the conserved density and spatial flux have the form
\begin{equation}
T = D_{\mathbf{x}}\cdot\mathbf{\Theta}= div \mathbf{\Theta}, \quad \mathbf{X} = D_{\mathbf{x}}\cdot\mathbf{\Psi}-D_t \mathbf{\Theta},
\label{triv_conlaw}
\end{equation}
whereby the continuity equations (\ref{conlaw}) and (\ref{integral}) hold as identities 
for some vector function $\mathbf{\Theta}$ and antisymmetric tensor function $\mathbf{\Psi}$ of $t,\mathbf{x},\rho,\mathbf{u}$, and $\mathbf x$-derivatives of $\rho$,$\mathbf{u}$.
The corresponding identity holding in a moving domain $V(t)$ takes the form
\begin{equation}
\frac{d}{dt}\int_{\p V(t)}\mathbf{\Theta}\cdot\hat{\mathbf{n}} d^{n-1}\sigma 
= -\int_{\p V(t)}{\boldsymbol\xi}\cdot\hat{\mathbf{n}} d^{n-1}\sigma 
= -\int_{V(t)}div {\boldsymbol\xi} d^n x
\label{mov_dom}
\end{equation}
where ${\boldsymbol\xi} = -D_t \mathbf{\Theta}-(D_\mathbf{x}\cdot\mathbf{\Theta)}\mathbf{u}$ is the spatial flux
through the moving boundary $\p V(t)$. As it stands, (\ref{mov_dom}) has no physical content.
However, an interesting observation is that if the moving-flux ${\boldsymbol\xi}$ is divergence-free for
all formal solutions of the Euler equations (\ref{veleqn})--(\ref{eos}) then the quantity $\int_{\p V(t)}\mathbf{\Theta}\cdot\hat{\mathbf{n}} d^{n-1}\sigma$ will be a nontrivial constant of motion.
Namely, 
when $\mathbf{u}(t,\mathbf{x})$ and $\rho(t,\mathbf{x})$ satisfy (\ref{veleqn})--(\ref{eos}), 
vector functions $\mathbf{\Theta}$ that satisfy the condition
\begin{equation}
0= -div {\boldsymbol\xi} = D_t div \mathbf{\Theta}+D_\mathbf{x}\cdot((div \mathbf{\Theta})\mathbf{u})= (D_t + \mathbf{u}\cdot D_\mathbf{x} + \nabla \cdot \mathbf{u})div\mathbf{\Theta}
\label{cond}
\end{equation}
lead to nontrivial conservation laws of the form
\begin{equation}
\frac{d}{dt}\int_{\p V(t)}\mathbf{\Theta}\cdot\hat{\mathbf{n}} d^{n-1}\sigma = 0
\label{nontri}
\end{equation}
for any boundary hypersurface $\p V(t)$ that is transported in the fluid.
In particular, it is sufficient for $\mathbf{\Theta}$ to satisfy the condition of vanishing flux
$-{\boldsymbol\xi} = D_t\mathbf{\Theta} + (D_\mathbf{x}\cdot\mathbf{\Theta)}\mathbf{u}=0$ for all formal solutions of (\ref{veleqn})--(\ref{eos}).
We will call (\ref{nontri}) a {\it moving-boundary} conservation law.

\subsection{Classification of kinematic conservation laws}

We first consider kinematic conservation laws as defined by the form 
\begin{equation}
T(t,\mathbf{x},\rho,\mathbf{u})
\label{mech}
\end{equation}
for the conserved density.
In the case of polytropic \esos/,
\begin{equation}
p=P(\rho)=\kappa\rho^\gamma,\quad \kappa,\gamma=const.
\label{polytropic}
\end{equation}
where pressure is proportional to a power of the density, 
all of the known local conservation laws (\ref {mech}) of the compressible polytropic Euler equations (\ref{veleqn}),(\ref{deneqn}),(\ref{polytropic}) 
in $n>1$ dimensions are summarized in the following table \cite{Ibr1973}.

\noindent \begin{tabular}{|c|c|c|c|}
\hline
Conserved density $T$   &  Description & Number  &  Exponent  \\
\hline
$\rho$ & Mass & $1$ & $\gamma$ arbitrary\\
$\rho\mathbf{u}$ & Momentum & $n$ & "\\
$\rho\mathbf{u}\wedge\mathbf{x}$ &Angular momentum & $n(n-1)/2$ & "\\
$\rho(t\mathbf{u}-\mathbf{x})$ & Galilean momentum & $n$ & "\\
$\frac{1}{2}\rho|\mathbf{u}|^2+\frac{\kappa}{\gamma-1}\rho^\gamma=E$ & Energy & $1$ & $\gamma \neq1$\\
$\rho(\frac{1}{2}|\mathbf{u}|^2+\kappa \ln \rho)$ & Energy & $1$ & $\gamma=1$\\
$tE-\frac{1}{2}\rho\mathbf{u}\cdot\mathbf{x}$ & Similarity energy & 1 & $\gamma=1+2/n$ \\
$t^2E-t\rho\mathbf{u}\cdot\mathbf{x}+\frac{1}{2}\rho|\mathbf{x}|^2$ & Dilational energy & 1 & "\\
\hline
\end{tabular}\\

We begin by stating a general classification of kinematic conservation laws with respect to all \esos/ (\ref{eos}).
Note that any conservation law of this form (\ref{mech}) is locally nontrivial since it does not contain $\mathbf x$-derivatives of $\rho$ or $\mathbf{u}$.

\textbf{Theorem 2.1:} 
(i) For a general \eos/ (\ref{eos}), the fluid conservation laws (\ref{mech}) in any dimension $n>1$ comprise a linear combination of
mass, momentum, angular momentum, Galilean momentum, and energy. 
In particular, for any spatial domain $V(t)\subset\mathbb R^n$ transported in the fluid:
\begin{eqnarray}
&&\frac{d}{dt}\int_{V(t)} \rho d^n x = 0, \label{kin_a} \\
&&\frac{d}{dt}\int_{V(t)} \rho\mathbf{u} d^nx = -\int_{\p V(t)}p\hat{\mathbf{n}}d^{n-1}\sigma, \label{kin_b} \\
&&\frac{d}{dt}\int_{V(t)} \rho\mathbf{u}\wedge\mathbf{x} d^nx = \int_{\p V(t)}p\mathbf{x}\wedge\hat{\mathbf{n}}d^{n-1}\sigma, \label{kin_c} \\
&&\frac{d}{dt}\int_{V(t)} \rho(t\mathbf{u}-\mathbf{x})d^nx = -\int_{\p V(t)}t p\hat{\mathbf{n}}d^{n-1}\sigma, \label{kin_d} \\
&&\frac{d}{dt}\int_{V(t)}(\tfrac{1}{2}\rho|\mathbf{u}|^2+\rho e)d^n x = -\int_{\p V(t)}p\mathbf{u}\cdot\hat{\mathbf{n}}d^{n-1}\sigma, \label{kin_e}
\end{eqnarray}
where 
\begin{equation}
p= P(\rho),\quad e=\int\rho^{-2}P(\rho)d\rho.
\label{energy}
\end{equation}

(ii) Modulo a constant shift in $p$, the only \eos/ for which extra conservation laws (\ref{mech}) arise is the polytropic case (\ref {polytropic}) with dimension-dependent exponent $\gamma=1+\frac{2}{n}$,
\begin{equation}
p=\kappa \rho^{1+\frac{2}{n}}, \quad \kappa = const.
\label{special}
\end{equation}
The admitted conservation laws consist of a linear combination of a similarity energy and a dilational energy. In particular:
\begin{eqnarray}
&&\frac{d}{dt}\int_{V(t)} (tE-\tfrac{1}{2}\rho\mathbf{u}\cdot\mathbf{x})d^nx = -\int_{\p V(t)}p(t\mathbf{u}-\tfrac{1}{2}\mathbf{x})\cdot\hat{\mathbf{n}} d^{n-1}\sigma, 
\label{kin_f} \\
&&\frac{d}{dt}\int_{V(t)} (t^2 E-t\rho\mathbf{u}\cdot\mathbf{x}+\tfrac{1}{2}\rho|\mathbf{x}|^2)d^n x = -\int_{\p V(t)}pt(t\mathbf{u}-\mathbf{x})\cdot\hat{\mathbf{n}} d^{n-1}\sigma, 
\label{kin_g}
\end{eqnarray}
where 
\begin{equation}
E=\tfrac{1}{2}\rho|\mathbf{u}|^2+\tfrac{1}{2}n\kappa\rho^{1+\frac{2}{n}} = \tfrac{1}{2}\rho|\mathbf{u}|^2+\tfrac{1}{2}np
\label{polyenergy}
\end{equation}
is the polytropic energy density.

The proof of this classification theorem is given in section \ref{proof2} using index notation. A summary of the conservation laws (\ref{kin_a})--(\ref{polyenergy}) written in explicit component form is presented in section 5. 

We remark that these conservation laws were first derived \cite{Ibr1973} for the case of irrotational fluid flow with a polytropic \eos/ (\ref{polytropic}). The Euler equations for such fluids in $n>1$ dimensions turn out to have a Lagrangian formulation  when a velocity potential is introduced (i.e., $\nabla \wedge \mathbf{u} = 0$ implies $\mathbf{u} = \nabla \Phi$), which allows local continuity equations to be classified in terms of point symmetries by means of Noether's Theorem \cite{BCA,Olv}. In particular mass conservation arises from invariance of the Euler-Lagrange fluid equations under shifts in the velocity potential. Invariance under space translations, rotations, Galilean boosts, and time translation respectively yields conservation of momentum, angular momentum, Galilean momentum, and energy.
For the special polytropic \eos/ (\ref{special}) the conserved similarity energy arises from a particular combination of scaling and dilation invariance that produces a variational symmetry, while the dilational energy corresponds to an extra symmetry \cite{Ovs} that is admitted only for this \eos/.

\subsection{Classification proof} \label{proof2}

The proof of Theorem 2.1 is based on explicitly solving the determining equations (\ref{deteqn}) by tensorial index methods.
We introduce the following index notation: $\mathbf{x}\leftrightarrow x^i$, $\mathbf{u}\leftrightarrow u^i$, $\nabla \rho \leftrightarrow \rho_{,i}$ and $\nabla \mathbf{u} \leftrightarrow u^j{}_{,i}$ 
(using a subscript comma  to denote partial derivatives), 
and $D_\mathbf{x} \leftrightarrow D_i$, where $i= 1,2,\ldots,n$; 
indices will be freely raised and lowered via the Kronecker symbols $\delta_{ij}$ and $\delta^{ij}$ (which are components of the Euclidean metric tensor and its inverse on $\mathbb R^n$ in Cartesian coordinates). 
The summation convention will apply to repeated indices.

In index notation,
\begin{equation}
u^i_t=-u^j u^i{}_{,j}-\rho^{-1}P'(\rho)\rho_,{}^i, \quad
\rho_t=-(\rho u^i)_{,i}, 
\label{eqn_11}
\end{equation}
are the Euler equations (\ref {veleqn})--(\ref {eos}). The spatial Euler operators with respect to $\rho$ and $u^i$ are given by  
\begin{equation}
E_\rho=\frac{\p}{\p \rho}-D_i\frac{\p}{\p \rho_{,i}}+ \cdots\quad \mbox{and}\quad E_{u^i}=\frac{\p}{\p u^i}-D_j\frac{\p}{\p u^i{}_{,j}}+ \cdots.
\end{equation}

For a conserved density of the kinematic form $T(t,x^i,\rho,u^i)$, we have
\begin{equation}
{\mathcal D}_t T= -T_\rho(u^i\rho_{,i}+\rho u^i{}_{,i})-T_{u^i}(u^ju^i{}_{,j}+\rho^{-1}P'(\rho)\rho_,{}^i)+T_t.
\label{tderiv}
\end{equation}
First applying the Euler operator $E_\rho$ to (\ref {tderiv}) we get
\begin{eqnarray*}
(\delta_{ij}\rho T_{\rho\rho}-\rho^{-1}P'(\rho) T_{u^iu^j})u^i{}_,{}^j+ \rho^{-1}P'(\rho) T_{u^ix_i}+T_{\rho x^i}u^i+T_{t\rho}
\end{eqnarray*}
which is a linear inhomogeneous scalar expression in $u^i{}_,{}^j$. Its coefficient must vanish, yielding the two equations
\begin{eqnarray}
&&\delta_{ij}\rho T_{\rho\rho}-\rho^{-1}P'(\rho) T_{u^iu^j}=0, \label{eqn1}\\
&&\rho^{-1}P'(\rho) T_{u^ix_i}+T_{\rho x^i}u^i+T_{t\rho}=0. \label{eqn2}
\end{eqnarray}
Next we apply the other Euler operator $E_{u^i}$ to (\ref {tderiv}), obtaining
\begin {eqnarray*}
(\delta_{ij}\rho T_{\rho\rho}-\rho^{-1}P'(\rho) T_{u^iu^j})\rho_,{}^j-(\rho T_{u^i\rho}-T_{u^i})u^j{}_{,j}+(\rho T_{u^j\rho}-T_{u^j})u^j{}_{,i}+ \rho T_{\rho x^i}+T_{u^ix^j}u^j+T_{tu^i}
\end{eqnarray*}
which is a linear inhomogeneous expression in $\rho_,{}^j$, $u^j{}_{,j}$ and $u^j{}_{,i}$. First we see the coefficient of $\rho_,{}^j$ yields the same terms as in (\ref {eqn1}).
Next, since $u^j{}_{,j}$ and $u^j{}_{,i}$ are linearly independent in $n>1$ dimensions, their coefficients must separately vanish, which yields
\begin{equation}
\rho T_{u^i\rho}-T_{u^i}=0. \label{eqn3}
\end{equation}
This leaves the inhomogeneous terms
\begin{equation}
\rho T_{\rho x^i}+T_{u^ix^j}u^j+T_{tu^i}=0. \label{eqn4}
\end{equation}

Hence the determining equations consist of (\ref {eqn1})--(\ref {eqn4}) to be solved for $T$.
We start from equation (\ref{eqn3}), which is first-order linear in $\rho$. By integrating with respect to $\rho$, and then doing a trivial integration with respect to $u^i$, we obtain 
\begin {equation}
T= \rho f(t,x^i,u^i)+g(t,x^i,\rho). \label{eqn5}
\end{equation}
Substituting (\ref{eqn5}) into (\ref{eqn1}) gives
\begin{equation}
\delta_{ij}\rho g_{\rho\rho} = P'(\rho)f_{u^iu^j} \label{eqn6}
\end{equation}
which separates with respect to $\rho,u^i$ into two equations
\begin{equation}
f_{u^ku^j}=c(t,x^i)\delta_{kj}, \quad
g_{\rho\rho}=c(t,x^i)\rho^{-1}P'(\rho), 
\label{eqn8}
\end{equation}
where $c(t,x^i)$ is a constant of separation.
Integration of (\ref{eqn8}) yields
\begin{eqnarray} 
&&f=c_1(t,x^i)+\tilde{c_j}(t,x^i)u^j+\tfrac{1}{2}c(t,x^i)u^ju_j, \label{eqn9}\\
&&g=c_0(t,x^i)+c_2(t,x^i)\rho+ c(t,x^i)\rho e, \label{eqn10}\\
&&e=\int\rho^{-2}P(\rho) d\rho, \label {e}
\end{eqnarray}
where $c_0$, $c_1$, $c_2$ and $\tilde{c_j}$ are constants of integration with respect to $\rho$ and $u^i$.
Thus, we have
\begin{equation}
T=c_0+\hat c\rho + \tilde{c_i}\rho u^i+c(\tfrac{1}{2}\rho u^iu_i+\rho e)
\label{T}
\end{equation}
where $\hat c=c_1+c_2$. Since the term $c_0$ is trivially conserved, we can put $c_0=0$.   

By substituting (\ref{T}) into (\ref {eqn2}) and (\ref{eqn4}), in each case we get a cubic polynomial in terms of $u^i$ whose separate coefficients must vanish. This leads to the following system of equations:
\begin{eqnarray}
&&c_{,i}=0, \label{eqni}\\
&&\tilde{c}_{j,i}+\tilde{c}_{i,j}+c_t\delta_{ij}=0, \label{eqnii}\\
&&\tilde{c_i}_t+\hat c_{,i}=0, \label{eqniii}\\
&&\hat c_t+P'(\rho)\tilde{c}_{i,}{}^i+(\rho e)'c_t=0. \label{eqniv}
\end{eqnarray}
To proceed, we note (\ref{eqni}) immediately implies $c=c(t)$.
Then (\ref{eqnii}) has the form of a time-dependent dilational Killing vector equation on $\tilde{c_i}$. To derive the solution, we first take the antisymmetrized derivative of (\ref{eqnii}), i.e. differentiating with respect to $x^k$ followed by antisymmetrizing in $j$ and $k$, which yields
\begin{equation}
(\tilde{c}_{j,k}-\tilde{c}_{k,j})_{,i}=0. \label{1na}
\end{equation}
Similarly, by taking the curl of (\ref{eqniii}), i.e. differentiating  with respect to $x^j$ and antisymmetrizing in $i$ and $j$, we obtain
\begin{equation}
(\tilde{c}_{i,j}-\tilde{c}_{j,i})_t=0. \label{1nb}
\end{equation}
Hence (\ref{1na}) and (\ref{1nb}) give
\begin{equation}
\tilde{c}_{i,j}-\tilde{c}_{j,i}=2C_{1ij} \label{1nc}
\end{equation}
where $C_{1ij}$ is an antisymmetric constant tensor.
Adding (\ref{1nc}) to (\ref{eqnii}), we get
\begin{equation}
2\tilde{c}_{i,j}=2C_{1ij}-c'(t)\delta_{ij} \label{1nd}
\end{equation}
and thus, by integration with respect to $x^j$, 
\begin{equation}
\tilde{c}_i = C_{0i}(t)+C_{1ij}x^j-\tfrac{1}{2}c'(t)x_i,\quad C_{1ij}=-C_{1ji}. \label{kv}
\end{equation}
Then (\ref{eqniii}) becomes
\begin{equation}
\hat c_{,i}=-C'_{0i}+\tfrac{1}{2}c''x_i
\label{cxi}
\end{equation}
which yields
\begin{equation}
\hat c=C_2(t)-C'_{0i}x^i+\tfrac{1}{4}c''x^ix_i.
\label{c}
\end{equation}
Finally, from (\ref {eqniv}), by using (\ref{c}) and the trace of (\ref{eqnii}), we get
\begin{equation}
C_2'-C''_{0i}x^i+\tfrac{1}{4}c'''x^ix_i+(-\tfrac{1}{2}nP'+(\rho e)')c'=0.
\label{eqiv}
\end{equation}
Splitting (\ref{eqiv}) with respect to $x^i$ yields 
\begin{eqnarray}
C''_{0i}=0,\quad c'''=0, \label{C0c}\\ 
C_2'+((\rho e)'-\tfrac{1}{2}nP')c'=0. \label{C2F}
\end{eqnarray}
From (\ref{C0c}) we have 
\begin{equation}
C_{0i}=a_{0i}+a_{1i}t,\quad c=b_0+b_1t+b_2t^2
\end{equation}
with constants $a_{0i}$, $a_{1i}$, $b_0$, $b_1$, $b_2$. Differentiating (\ref{C2F}) with respect to $\rho$ gives 
\begin{equation}
((\rho e)'-\tfrac{1}{2}nP')'c'=0 \label{C2Fr}
\end{equation}
which leads to the following two cases.

\textbf{Case} $c'=0$: Hence $c=b_0$ and $b_1=b_2=0$, which implies $C_2=const.$ from (\ref{C2F}). Then (\ref {kv}) and (\ref{c}) yield
\begin{equation}
\tilde{c_i} = a_{0i}+a_{1i}t+C_{1ij}x^j,\quad \hat c=C_2-a_{1i}x^i, 
\end{equation}
whence from (\ref{T}) we obtain
\begin{eqnarray*}
T=C_2\rho+a_{0i}\rho u^i + a_{1i}\rho (t u^i-x^i)+C_{1ij}\rho u^ix^j+b_0\rho (\tfrac{1}{2}u^iu_i+e)
\end{eqnarray*}
where $C_2$, $C_{1ij}=-C_{1ji}$, $a_{0i}$, $a_{1i}$, $b_0$ are arbitrary constants and $e$ is a function of $\rho$ given by (\ref{e}). 

\textbf{Case} $(\rho e)''-\frac{n}{2}P''=0$: Simplifying  $(\rho e)''=\rho e''+2e'=\rho^{-1}P'$ through (\ref {e}), we get a linear ODE
\begin{equation}
n\rho P''-2P'=0
\label {ode}
\end{equation}
 which has the general solution $P(\rho)=d_0+d_1\rho^{1+2/n}$, where $d_0$, $d_1$ are constants. Note that we can put $d_0=0$ since the pressure $p=P(\rho)$ can be shifted (without loss of generality) by an arbitrary constant. This implies   
\begin{equation}
P=d_1\rho^{1+2/n},\quad e = \tfrac{1}{2}nd_1\rho^{2/n}+d_2
\end{equation}
from (\ref{e}), with a constant of integration $d_2$. Thus, (\ref {C2F}) becomes $C_2'+d_2c'=0$ whence
\begin{equation}
C_2=b_3-d_2c(t)=b_3-d_2b_0-d_2b_1t-d_2b_2t^2
\label{C_2}
\end{equation}
where $b_3$ is an integration constant. Then (\ref {kv}) and (\ref{c}) yield
\begin{equation}
\tilde{c_i} = a_{0i}+a_{1i}t+C_{1ij}x^j-\tfrac{1}{2}(b_1+2 b_2t)x_i, \quad \hat c=C_2(t)-a_{1i}x^i+\tfrac{1}{2}b_2x^ix_i.
\label{cc}
\end{equation}
Substituting (\ref{C_2}) and (\ref{cc}) into (\ref{T}), we find that the terms involving $d_2 c(t)$ cancel out, giving  
\begin{eqnarray*} 
T &=&
b_3\rho+a_{0i}\rho u^i + a_{1i}\rho (t u^i-x^i)+C_{1ij}\rho u^ix^j+b_0\rho(\tfrac{1}{2} u^iu_i+ \tilde{e}) 
\\&&
+b_1\rho (\tfrac{1}{2} u^i(tu_i-x_i)+t \tilde{e}) + b_2\rho(\tfrac{1}{2} (tu^i-x^i)(tu_i-x_i)+t^2 \tilde{e})
\end{eqnarray*}
with 
\begin{equation}
\tilde{e} = \tfrac{1}{2}nP/\rho, \quad P = d_1 \rho^{1+n/2}
\end{equation}
where $d_1$, $b_0$, $b_1$, $b_2$, $b_3$, $a_{0i}$, $a_{1i}$, $C_{1ij}=-C_{1ji}$ are arbitrary constants.

This completes the proof of Theorem 2.1.

\section {Vorticity conservation laws}

In $n>1$ dimensions, the curl of the fluid velocity is the antisymmetric tensor
\begin{equation}
{\boldsymbol\omega}=\nabla\wedge\mathbf{u} 
\label {curl}
\end{equation}
which satisfies the identities $\nabla\wedge{\boldsymbol\omega}=0$ and $\nabla\cdot{\boldsymbol\omega}=\Delta\mathbf{u}-\nabla(\nabla\cdot\mathbf{u})$.
There is a natural odd-parity expression that can be constructed purely out of products of ${\boldsymbol\omega}$ and the spatial orientation tensor ${\boldsymbol\epsilon}$ as follows.
(Recall, ${\boldsymbol\epsilon}$ is a rank $n$ totally skew-symmetric tensor whose components in Cartesian coordinates for $\mathbb R^n$ are given by the Levi-Civita symbol. 
In particular, up to a choice of sign, ${\boldsymbol\epsilon}$ is determined by its properties $\nabla{\boldsymbol\epsilon}=0$ and $|{\boldsymbol\epsilon}|^2={\boldsymbol\epsilon}\cdot{\boldsymbol\epsilon}=n!$.)

When the spatial dimension is even, say $n=2m$,
\begin{equation}
\varpi ={\boldsymbol\epsilon}\cdot(\underbrace{(\nabla\wedge\mathbf{u})\wedge\cdots\wedge(\nabla\wedge\mathbf{u})}_{\textrm{$m$  times}})
\label {even}
\end{equation} 
defines a vorticity scalar that has odd parity since ${\boldsymbol\epsilon}$ changes sign under spatial reflections. Similarly, when the spatial dimension is odd, say $n=2m+1$, the analogous expression
\begin{equation}
{\boldsymbol\varpi} ={\boldsymbol\epsilon}\cdot(\underbrace{(\nabla\wedge\mathbf{u})\wedge\cdots\wedge(\nabla\wedge\mathbf{u})}_{\textrm{$m$  times}})
\label {odd}
\end{equation}
defines a vorticity vector with odd parity under spatial reflections.
These expressions (\ref{even}) and (\ref{odd}) can be written in the more compact notation $\ast({\boldsymbol\omega^m})$, where $\ast$ is the Hodge dual operator (acting by contraction with respect to ${\boldsymbol\epsilon}$) and $m=[n/2]$ is a positive integer.

We now define vorticity conservation laws to have the form 
\begin{equation}
T(\rho,\mathbf{u},\ast({\boldsymbol\omega^m}))
\label{vorticity}
\end{equation}
for the conserved density, such that the expression (\ref{vorticity}) possesses odd parity under spatial reflections as follows. 
Let ${\mathcal P}_{\boldsymbol{\ell}}$ be a reflection operator defined with respect to a spatial unit vector ${\boldsymbol{\ell}}$ in $\mathbb R^n$.
Specifically, ${\mathcal P}_{\boldsymbol{\ell}}$ reverses all vectors parallel to ${\boldsymbol{\ell}}$ while leaving invariant all vectors in the hyperplane orthogonal to ${\boldsymbol{\ell}}$ through the origin.
Note ${\mathcal P}_{\boldsymbol{\ell}}$ extends to act on tensors by multi-linearity and acts as the identity on scalars.
Then for any choice of ${\boldsymbol{\ell}}$, ${\mathcal P}_{\boldsymbol{\ell}}$ satisfies the properties ${\mathcal P}_{\boldsymbol{\ell}}{\boldsymbol\epsilon}=-{\boldsymbol\epsilon}$ and ${\mathcal P}_{\boldsymbol{\ell}}^2=1$, so consequently the parity of a conserved density (\ref{vorticity}) will be odd under spatial reflections iff
\begin{equation}
T({\mathcal P}\rho, {\mathcal P}\mathbf{u}, \ast({\mathcal P}{\boldsymbol\omega^m}))=-{\mathcal P}T(\rho,\mathbf{u},\ast({\boldsymbol\omega^m})). \label{vort1}
\end{equation}
Note that, in contrast, the parity of all the kinematic conserved densities (\ref{kin_a})--(\ref{kin_g}) is even,
\begin{equation}
T(t,{\mathcal P}\mathbf{x},{\mathcal P}\rho, {\mathcal P}\mathbf{u})={\mathcal P}T(t,\mathbf{x},\rho,\mathbf{u}).
\label{vort2}
\end{equation}

Our main result will now be a complete classification of vorticity conservation laws for all \esos/ (\ref{eos}).

\textbf{Theorem 3.1:}
For any \eos/ (\ref{eos}) the only nontrivial fluid conservation laws (\ref {vorticity})--(\ref{vort1}) in dimensions $n>1$ are given by helicity
\begin{equation}
\frac{d}{dt}\int_{V(t)} \mathbf{u}\cdot {\boldsymbol\varpi}d^n x = -\int_{\p V(t)}(f(\rho)-\tfrac{1}{2}|\mathbf{u}|^2){\boldsymbol\varpi}\cdot\hat{\mathbf{n}}d^{n-1}\sigma\\
\label{vconlaw1}
\end{equation}
for odd dimensions $n=2m+1$, where 
\begin{equation}
f(\rho)=\int\rho^{-1}P'(\rho)d\rho = \frac{P}{\rho}+e, \label{heli}
\end{equation}
and generalized enstrophy
\begin{equation}
\frac{d}{dt}\int_{V(t)} \rho f(\varpi/\rho)d^n x = 0\\
\label{vconlaw2}
\end{equation} 
for even dimensions $n=2m$, where $f$ is any nonlinear odd function of $\varpi/\rho$.
In particular, there are no special \esos/ that admit extra vorticity conservation laws.

The helicity and enstrophy conservation laws were first derived by means of a Hamiltonian Casimir analysis \cite{KC,AK}, as we will discuss further in section \ref{corresp}, which is a more restrictive analysis than directly solving the determining equations (\ref{deteqn}) for conserved densities. 
By comparison, our classification has more generality and actually holds without imposing the odd-parity condition (\ref{vort1}) if we consider conserved densities (\ref{vorticity}) that just have an essential dependence on ${\boldsymbol\omega}$.
The proof is given in section 3.1 and a summary of the conservation laws (\ref{vconlaw1})--(\ref{vconlaw2}) in explicit component form is shown in section 5.

The local continuity equations underlying these conservation laws (\ref{vconlaw1})--(\ref{vconlaw2}) are readily obtained from the transport equations satisfied by ${\boldsymbol\varpi}$ and $\varpi$. 
Through (\ref {curl}), we first note the identity 
\begin{eqnarray*}
\mathbf{u}\cdot{\boldsymbol\omega} = \mathbf{u}\cdot\nabla \mathbf{u}-\tfrac{1}{2}\nabla(\mathbf{u}\cdot\mathbf{u})
\end{eqnarray*}
whence the Euler equation (\ref{veleqn}) for the fluid velocity  can be written in the form 
\begin{equation}
\p_t \mathbf{u}+\mathbf{u}\cdot{\boldsymbol\omega}+\nabla(\tfrac{1}{2}|\mathbf{u}|^2+f(\rho)) =0 \label{ut}
\end{equation}
where $f(\rho)$ is given by (\ref{heli}) through the \eos/ (\ref{eos}). 
Taking the curl of (\ref{ut}) we get 
\begin{equation}
\p_t{\boldsymbol\omega} = \nabla\wedge ({\boldsymbol\omega}\cdot\mathbf{u}).
\label {omegat}
\end{equation}
Substitution of (\ref{omegat}) into the time derivatives of (\ref{even}) and (\ref{odd}) then yields the respective transport equations
\begin{equation}
\p_t\varpi+\mathbf{u}\cdot\nabla\varpi = -(\nabla\cdot\mathbf{u})\varpi,\quad
\p_t{\boldsymbol\varpi}+\mathbf{u}\cdot\nabla{\boldsymbol\varpi} = {\boldsymbol\varpi}\cdot\nabla\mathbf{u}-(\nabla\cdot\mathbf{u}){\boldsymbol\varpi},
\end{equation}
or equivalently,
\begin{eqnarray}
&&\p_t\varpi = -\nabla\cdot(\varpi\mathbf{u}), \label{tr1}\\
&&\p_t{\boldsymbol\varpi} = \nabla\cdot({\boldsymbol\varpi}\wedge\mathbf{u}). \label{tr2}
\end{eqnarray}

The vorticity transport equations (\ref{tr1}) and (\ref{tr2}) each have the form of a
local continuity equation holding in even ($n=2m$) and odd ($n=2m+1$) dimensions, with respective conserved densities
$T=\varpi$ and $T={\boldsymbol\varpi}$.
As we will now show, both of these vorticity conservation laws are locally trivial and therefore fall outside of our classification theorem. 
From the explicit expressions (\ref{even}) and (\ref{odd}) for the vorticity densities in terms of the curl of the fluid velocity
${\boldsymbol\omega} = \nabla\wedge \mathbf{u}$, we find that in both even and odd dimensions
\begin{equation}
T=\ast({\boldsymbol\omega}^m)=\nabla \cdot \ast(\mathbf{u}\wedge {\boldsymbol\omega}^{m-1}) = div\mathbf{\Theta}
\label{vorT}
\end{equation}
is a spatial divergence.
The fluid equations (\ref{ut}) and (\ref{omegat}) yield the time derivative of $\mathbf{\Theta}=\ast(\mathbf{u}\wedge {\boldsymbol\omega}^{m-1})$ to be
\begin{equation}
\p_t \mathbf{\Theta}= -\mathbf{u}\wedge\ast({\boldsymbol\omega}^m)-\nabla \cdot\mathbf{\Psi}
\label{thetat}
\end{equation}
where $\mathbf{\Psi}=(\frac{1}{2}|\mathbf{u}|^2-f(\rho))\ast({\boldsymbol\omega}^{m-1})+\mathbf{u}\wedge(\mathbf{u}\cdot\ast({\boldsymbol\omega}^{m-1}))$ is an antisymmetric tensor.
Since in both even and odd dimensions the expression
\begin{equation}
\mathbf{X} = \mathbf{u}\wedge\ast({\boldsymbol\omega}^m)
\label{X}
\end{equation}
is the spatial flux,
$\mathbf{X} = \varpi \mathbf{u}$ and $\mathbf{X} =\mathbf{u} \wedge {\boldsymbol\varpi}$,
arising from the vorticity transport equations (\ref{tr1}) and (\ref{tr2}),
we then see (\ref{X}) and (\ref{vorT}) have the form (\ref{triv_conlaw}) of a locally trivial spatial flux and a locally trivial conserved density.

It is interesting to investigate the corresponding vorticity flux ${\boldsymbol \xi}=\mathbf{X}-T\mathbf{u}$ through a moving domain boundary in the fluid.

In odd dimensions $n=2m+1$, we can write
\begin{equation}
{\boldsymbol\varpi} = -\nabla \cdot(\mathbf{u}\cdot\mathbf{W})
\end{equation}
where 
\begin{equation}
\mathbf{W}=\ast({\boldsymbol\omega}^{m-1})=\tfrac{1}{m}\p_{\boldsymbol\omega}{\boldsymbol\varpi}
\end{equation}
is a totally skew-symmetric tensor of rank $3$. Hence for any domain $V(t)$ transported in the fluid, we obtain
\begin{equation}
-\frac{d}{dt}\int_{V(t)} {\boldsymbol\varpi} d^{2m+1} x 
= \frac{d}{dt}\int_{\p V(t)} \mathbf{W}\cdot(\mathbf{u}\wedge \hat{\mathbf{n}})d^{2m}\sigma 
= \int_{\p V(t)}\mathbf{u}({\boldsymbol\varpi}\cdot \hat{\mathbf{n}}) d^{2m}\sigma
\label{eqn80}
\end{equation}
for all formal solutions of the dynamical equations (\ref{ut}) and (\ref{tr2}). 
Since the moving-flux through $\p V(t)$ fails to vanish, (\ref{eqn80}) does not yield a constant of motion.

For even dimensions $n=2m$, we have
\begin{equation}
\varpi = -\nabla \cdot(\mathbf{u}\cdot\mathbf{w}) = \nabla \cdot(\mathbf{w}\cdot\mathbf{u})
\end{equation}
where
\begin{equation}
\mathbf{w}= \ast({\boldsymbol\omega}^{m-1}) =\tfrac{1}{m} \p_{\boldsymbol\omega} \varpi
\end{equation}
is an antisymmetric tensor.
Then for all formal solutions of the dynamical equations (\ref{ut}) and (\ref{tr1}), we obtain
\begin{equation}
-\frac{d}{dt}\int_{V(t)} \varpi d^{2m} x = \frac{d}{dt}\int_{\p V(t)} \mathbf{w} \cdot(\mathbf{u}\wedge \hat{\mathbf{n}})d^{2m-1}\sigma = 0.
\label{eqn83}
\end{equation}
Thus (\ref{eqn83}) yields a constant of motion for the moving boundary $\p V(t)$ in $\mathbb R^{2m}$.
It describes an even-dimensional generalization of Kelvin's circulation for isentropic fluid flow in two dimensions \cite{Batch}. 
Specifically, when $n=2$ $(m=1)$, the vorticity scalar is given by
\begin{equation}
\varpi = {\boldsymbol\epsilon}\cdot{\boldsymbol\omega}=\nabla \cdot(\ast \mathbf{u})
\label{eqn84}
\end{equation}
and hence $\mathbf{w}={\boldsymbol\epsilon}$ is the spatial orientation tensor.
This yields $\mathbf{w}\cdot(\mathbf{u}\wedge \hat{\mathbf{n}}) = \mathbf{u}\cdot\ast\hat{\mathbf{n}}$, and then the conservation law (\ref{eqn83}) becomes Helmholtz's circulation theorem \cite{chorin} 
\begin{equation}
-\frac{d}{dt}\int_{V(t)} \varpi d^{2} x = \frac{d}{dt}\int_{\p V(t)} \mathbf{u} \cdot\ast\hat{\mathbf{n}}d\sigma = 0
\label{kelvin}
\end{equation}
where $\p V(t)$ is a closed curve in $\mathbb R^2$, $\ast\hat{\mathbf{n}}$ is a unit tangent vector along $\p V(t)$ and
$d\sigma$ is the arclength element.
By writing $d\mathbf{s}= \ast\hat{\mathbf{n}}d\sigma$ we see (\ref{kelvin}) states that the line integral
\begin{equation}
\oint \mathbf{u} \cdot d\mathbf{s}
\label{line}
\end{equation}
defining the circulation of the
fluid velocity around a curve transported in the fluid is a constant of the fluid motion.
For higher dimensions $n=2m$ $(m>1)$, we can write the moving-boundary conservation law (\ref{eqn83}) in an analogous form by noting
\begin{equation}
\mathbf{w}\cdot(\mathbf{u} \wedge \hat{\mathbf{n}}) = (\mathbf{u}\wedge{\boldsymbol\omega}^{m-1})\cdot\ast\hat{\mathbf{n}}
\label{analog}
\end{equation}
where $\ast\hat{\mathbf{n}}$ is the volume tensor for the boundary hypersurface $\p V(t)$ in $\mathbb R^{2m}$.
Then (\ref{eqn83}) becomes
\begin{equation}
-\frac{d}{dt}\int_{V(t)} \varpi d^{2m} x = \frac{d}{dt}\int_{\p V(t)}(\mathbf{u}\wedge{\boldsymbol\omega}^{m-1})\cdot\ast\hat{\mathbf{n}}d^{2m-1}\sigma = 0
\label{eqn88}
\end{equation}
which states that the (hyper)surface integral
\begin{equation}
\oint_{\p V(t)}(\mathbf{u}\wedge{\boldsymbol\omega}^{m-1})\cdot d\mathbf{A}
\label{const}
\end{equation}
is a constant of the fluid motion, 
where $d\mathbf{A}= \ast\hat{\mathbf{n}}d^{2m-1}\sigma$ denotes the volume element
for the moving-boundary (hyper)surface $\p V(t)$ in even dimensions $n=2m$
analogous to $d\mathbf{s}=\ast\hat{\mathbf{n}}d\sigma$ for moving-boundary curves in two dimensions.

\textbf{Proposition 3.2:} 
The only moving-boundary conservation law (\ref{nontri}) of vorticity type is the generalized circulation (\ref{eqn88}) holding for any \eos/ in all even dimensions.

The proof of this classification is given in the next section. 

\subsection{Classification proof}

We will use the same index notation introduced in section 2.2 for the proof of Theorem 2.1.
To begin, we write out the component form of the fluid curl, the vorticity scalar and vector, along with their transport equations:
\begin{eqnarray*}
&&{\boldsymbol\omega} \leftrightarrow \omega^{ij} = \tfrac{1}{2}(u^i{}_,{}^j-u^j{}_,{}^i) = u^{[i}{}_,{}^{j]},\\
&&\varpi \leftrightarrow \varpi =\epsilon^{i_1j_1\cdots i_mj_m}\omega_{i_1j_1}\cdots\omega_{i_mj_m},\quad m=n/2,\\
&&{\boldsymbol\varpi} \leftrightarrow \varpi^i = \epsilon^{i j_1k_1\cdots j_mk_m}\omega_{j_1k_1}\cdots\omega_{j_mk_m}, \quad m=(n-1)/2, \\ 
&&\omega^{ij}_t = (u_k\omega^{ki})_,{}^j-(u_k\omega^{kj})_,{}^i = u_{k,}{}^i \omega^{jk} - u_{k,}{}^j \omega^{ik} - u^k \omega^{ij}{}_{,k},\\
&&\varpi_t = -(\varpi u^i){}_{,i}= -\varpi u^i{}_{,i}- u^i \varpi_{,i}, \\
&&\varpi^i_t = (\varpi^j u^i-\varpi^i u^j)_{,j}= \varpi^j u^i{}_{,j}- \varpi^i u^j{}_{,j}-u^j\varpi^i{}_{,j}.
\end{eqnarray*}
In addition, we will need the component form for
\begin{eqnarray*}
&&\mathbf{w} \leftrightarrow w^{ij} =\epsilon^{iji_1j_1\cdots i_{m-1}j_{m-1}}\omega_{i_1j_1}\cdots\omega_{i_{m-1}j_{m-1}},\quad m=n/2,\\
&&\mathbf{W} \leftrightarrow W^{ijk} = \epsilon^{ijk j_1k_1\cdots j_{m-1}k_{m-1}}\omega_{j_1k_1}\cdots\omega_{j_{m-1}k_{m-1}}, \quad m=(n-1)/2, \\
&& \nabla\varpi \leftrightarrow \varpi_{,i} = w^{jk} \omega_{jk}{}_{,i}, \quad
\nabla{\boldsymbol\varpi} \leftrightarrow \varpi^j{}_{,i} = W^{jkl} \omega_{kl}{}_{,i}.
\end{eqnarray*}
Here $\epsilon^{i_1\cdots i_n} = \epsilon^{[i_1\cdots i_n]}\leftrightarrow {\boldsymbol\epsilon}$, $\delta_{ij} = \delta_{(ij)} \leftrightarrow \mathbf{g}$
are the components of the spatial orientation tensor and the Euclidean metric tensor; round brackets denote symmetrization of the enclosed indices, and square brackets denote antisymmetrization.

The proof of Theorem 3.1 and Proposition 3.2 proceeds by explicitly solving the determining equations (\ref{deteqn})
for conserved densities of vorticity type in even and odd dimensions $n>1$.
Recall, the Euler equations (\ref{veleqn})--(\ref{eos}) are given by
\begin{equation}
u^i_t=-u^j u^i{}_{,j}-f'(\rho)\rho_,{}^i, \quad
\rho_t=-(\rho u^i)_{,i}, \label{eqn_2}
\end{equation}
where
\begin{equation}
f'(\rho)=P'(\rho)/\rho \not\equiv 0.
\end{equation}

\textbf{Case} $n=2m$: For a conserved density of the form $T(\rho,u^i,\varpi)$, its time derivative is given by 
\begin{eqnarray}
-{\mathcal D}_t T &=& (\rho u^i)_{,i}T_\rho +(u^ju^i{}_{,j}+\rho_,{}^i f')T_{u^i}+(\varpi u^i)_{,i}T_{\varpi} \nonumber \\
&=& u^i(\rho_{,i} T_\rho +u^j{}_{,i} T_{u^j}+ \varpi_{,i}T_\varpi) + u^i{}_{,i}(\rho T_\rho+\varpi T_\varpi) +\rho_,{}^i f' T_{u^i}.
\label{eqn_a1}
\end{eqnarray} 
In (\ref{eqn_a1}), we see the coefficient of $u^i$ equals $D_iT$ by the chain rule, which allows us to write the corresponding terms as $u^iD_iT=D_i(u^iT)-u^i{}_{,i}T$. Hence 
\begin{equation}
{\mathcal D}_t T= D_i(-u^i T)+u^i{}_{,i}A- \rho_,{}^i f' T_{u^i}
\label{eqn_b1}
\end{equation}
where
\begin{equation}
A= T- \rho T_\rho - \varpi T_\varpi.
\label{eqn_c1}
\end{equation}

To begin we substitute (\ref{eqn_b1}) into the first determining equation 
\begin{equation}
0=E_\rho({\mathcal D}_t T) = \tr{u} A_\rho + u^{ij}T_{u^i u^j}+\varpi_,{}^i T_{u^i\varpi}
\label{eqn_d1}
\end{equation}
where we have introduced the notation
\begin{equation}
\tr{u}=\delta_{ij} u^{ij}=u^i{}_{,i},\quad u^{ij}=\tfrac{1}{2}(u^i{}_,{}^j+u^j{}_,{}^i)=  u^{(i}{}_,{}^{j)}.
\label{eqn_not}
\end{equation}
Since $T$ does not contain any derivatives of $\varpi$, the coefficient of $\varpi_,{}^i$ in (\ref{eqn_d1}) must vanish,
\begin{equation}
T_{u^i \varpi} = 0.
\label{eqn_e1}
\end{equation}
Integrating (\ref{eqn_e1}) and dropping a kinematic term $c(\rho,u^i)$ that does not involve $\varpi$, we get
\begin{equation}
T = a(\rho,\varpi).
\label{eqn_f1}
\end{equation}
Then  (\ref{eqn_d1}) reduces to $0 = \tr{u} A_\rho$ and hence we obtain 
\begin{equation}
0 = A_\rho = -\rho a_{\rho \rho} -\varpi a_{\varpi \rho}.
\label{eqn_g1}
\end{equation}
This is a first-order linear PDE for $a_\rho$, which yields
\begin{equation}
a = b(\varpi) + \rho c(\varpi/\rho).
\label{eqn_h1}
\end{equation}

Thus from (\ref{eqn_f1}) and (\ref{eqn_h1}) we have
\begin{equation}
T = b(\varpi) + \rho c(\varpi/\rho)
\label{eqn_i10}
\end{equation}
and
\begin{equation}
{\mathcal D}_t T = D_i(-u^i T) + u^i{}_{,i} (b-\varpi b').
\label{eqn_j1}
\end{equation}
Now we substitute (\ref{eqn_j1}) into the second determining equation
\begin{equation}
0=E_{u^i}({\mathcal D}_t T) = \varpi_{,i}B+m\tr{u}_,{}^j \varpi_{ji}B + m\varpi_,{}^j w_{ji}\tr{u}B'
\label{eqn_l10}
\end{equation}
where
\begin{equation}
B=\varpi b''.
\label{eqn_l2}
\end{equation}
Note that $b(\varpi)$ has no dependence on $\tr{u}$ since $\varpi$ does not contain $u^{ij}$ (which is linearly independent of $\varpi_{ij}$). Consequently, the coefficients of all terms in (\ref{eqn_l10}) must vanish, yielding $B=B'=0$ so thus $b''=0$.
Hence, we obtain
\begin{equation}
b=b_0+b_1\varpi
\label{eqn_m1}
\end{equation}
with constants $b_0$, $b_1$.
Therefore, (\ref{eqn_m1}) and (\ref{eqn_i10}) give the result
\begin{equation}
T = b_1\varpi + \rho c(\varpi/\rho) + b_0
\label{eqn_n1}
\end{equation}
with
\begin{equation}
{\mathcal D}_t T = -D_i(u^i T)
\label{eqn_o1}
\end{equation}
from (\ref{eqn_j1}), where $b_1$, $b_0$ are arbitrary constants and $c$ is an arbitrary function of $\varpi/\rho$.
Since the term $b_0$ is trivially conserved, we can put $b_0=0$.

We now note that the first term in (\ref{eqn_n1}) is a trivial conserved density
\begin{equation}
b_1\varpi =b_1(\varpi^{ij} u_j)_{,i} = D_i(b_1\varpi^{ij} u_j).
\label{eqn_p1}
\end{equation}
Its corresponding moving-flux vanishes, $\xi^i = 0$, due to the form of (\ref{eqn_o1}).
Similarly, if $c=\tilde{c}\varpi/\rho$ is a linear function of $\varpi/\rho$, 
where $\tilde{c}$ is a constant, 
then the second term in (\ref{eqn_n1}) becomes a trivial conserved density, $\rho c(\varpi/\rho) = \tilde{c} \varpi$.
This proves Theorem 3.1 and Proposition 3.2 in the even-dimensional case.

\textbf{Case} $n=2m+1$: Here we will need the identities
\begin{eqnarray}
&&W^{i(jk)} = W^{(ij)k}=0, \label{id3} \\
&&\varpi^i{}_{,i}=0, \label{id1}\\ 
&&\varpi^i \omega_{ij}=0, \label{id2}
\end{eqnarray}
which are consequences of
\begin{eqnarray}
&& W^{ijk}=W^{[ijk]}, \label{prop1}\\
&& W^{ijk}\omega_{ij}{}_{,k}=W^{ijk} u_{j,ik}=0, \label{prop2}\\
&&\omega_{jk}\varpi^k \epsilon_{ij_1k_1\cdots j_mk_m} = n!\omega_{j[i}\omega_{j_1k_1} \cdots \omega_{j_mk_m]}=n! \omega_{[ji}\omega_{j_1k_1} \cdots \omega_{j_mk_m]} = 0, \label{prop3}
\end{eqnarray}
where (\ref{prop3}) follows from the fact that there are no totally skew-symmetric tensors of rank $2m+2>n=2m+1$. 

Now for a conserved density $T(\rho,u^i,\varpi^i)$, by the same steps followed in the previous case to evaluate the time derivative, we have
\begin{equation}
{\mathcal D}_t T= D_i(-u^i T)+u^i{}_{,i}(T-\rho T_\rho - \varpi^j T_{\varpi^j})+ u^i{}_{,j}\varpi^j T_{\varpi^i}-\rho_,{}^i
f'T_{u^i}.
\label{eqn_b}
\end{equation}

To proceed we substitute (\ref{eqn_b}) into the first determining equation $0=E_\rho({\mathcal D}_t T)$. This yields the terms
\begin{equation}
0=\omega^{ij}\varpi_j T_{\rho \varpi^i}-\tr{u}(\rho T_{\rho \rho}+\varpi^i T_{\rho \varpi^i}) + u^{ij}(\varpi_j T_{\rho \varpi^i}+f' T_{u^iu^j})
+ \varpi^j{}_,{}^i f'T_{u^i\varpi^j}
\label{eqn_d}
\end{equation}
using the notation (\ref{eqn_not}).
To start, we observe the first term in (\ref{eqn_d}) vanishes by the identity (\ref{id2}). 
Next, since $T$ does not contain any derivatives of $\varpi^i$, the last term in (\ref{eqn_d}) must vanish modulo the identity (\ref{id1}). This implies 
\begin{equation}
T_{u^i \varpi^j} = a(\rho,u^k,\varpi^k)\delta_{ij}.
\label{eqn_e}
\end{equation}
Applying the derivative operator $\p_{u^k}$ to (\ref{eqn_e}) and antisymmetrizing in $[jk]$, we get
\begin{equation}
\delta_{ij}a_{u^k}-\delta_{ik}a_{u^j} = 0.
\label{deriv}
\end{equation}
The trace of this equation with respect to $(ij)$ yields $(n-1)a_{u^k} = 0$, whence in $n>1$ dimensions we obtain
\begin{equation}
a_{u^k} = 0.
\label{eqn_f}
\end{equation}
By also applying the derivative operator $\p_{\varpi^k}$ to (\ref{eqn_e}), we similarly get
\begin{equation}
a_{\varpi^k} = 0.
\label{eqn_g}
\end{equation}
Integration of (\ref{eqn_e}), (\ref{eqn_f}), (\ref{eqn_g}) then yields
\begin{equation}
T = a(\rho)u^i\varpi_i + b(\rho,\varpi^i).
\label{eqn_h}
\end{equation}
Here we have dropped an integration constant $c(\rho,u^i)$ since it does not involve $\varpi^i$ (i.e. it is of kinematic form).
Hence the determining equation (\ref{eqn_d}) reduces to
\begin{equation}
0 = u^{ij}(B_{ij}-\delta_{ij}A)
\label{eqn_i1}
\end{equation}
with coefficients
\begin{eqnarray}
&&A = \rho b_{\rho \rho} + \varpi^i b_{i \rho}, \label{j1}\\
&&B_{ij} = \varpi_{(j} b_{i)}\rho + a' \varpi_{(j} u_{i)}, \label{j2}
\end{eqnarray}
in terms of the notation $b_i = b_{\varpi^i}$. 
Since $\varpi^i$ does not contain $u^{ij}$ we see that $b(\rho,\varpi^i)$ has no dependence on $u^{ij}$ (or $\tr{u}$). 
Consequently, (\ref{eqn_i1}) implies that the coefficient of $u^{ij}$ must vanish,
\begin{equation}
B_{ij} = A\delta_{ij}.
\label{eqn_k}
\end{equation}
By considering the product of (\ref{j2}) with $\varpi_k \varpi_l$ and antisymmetrizing in $[il]$ and $[jk]$, we find
$\varpi_{[l} B_{i][j}\varpi_{k]} = 0$.
The same antisymmetric product applied to (\ref{eqn_k}) then implies $A \varpi_{[l}\delta_{i][j}\varpi_{k]} = 0$
which gives
\begin{equation}
A=0
\label{a0}
\end{equation}
and hence
\begin{equation}
B_{ij}=0.
\label{bij}
\end{equation}
Now by taking the product of (\ref{bij}) with $\varpi_k$ antisymmetrized in $[jk]$, we get
\begin{equation}
\varpi_{[k} b_{j]\rho} + a'\varpi_{[k} u_{j]} = 0.
\label{eqn_i2}
\end{equation}
This can hold only if $a' =0$ and $b_{j \rho} = \varpi_j c(\rho,\varpi^i)$.
Then (\ref{j2}) becomes
\begin{equation}
B_{ij} = c\varpi_i \varpi_j
\label{eqn_i3}
\end{equation}
whence (\ref{bij}) implies $c=0$ and so $b_{j\rho} = 0$.
Thus we have
\begin{equation}
b_{\varpi^i \rho} = 0
\label{eqn_l}
\end{equation}
whose solution is $b = \tilde{b}(\varpi^i) +\tilde{c}(\rho)$.
Since $\tilde{c}(\rho)$ contributes only a kinematic term in (\ref{eqn_h}), it will be dropped hereafter.
Then 
\begin{equation}
b=\tilde{b}(\varpi^i), \quad a=const.
\label{eqn_l1}
\end{equation}
satisfies both (\ref{bij}) and (\ref{a0}).

Thus we have, from (\ref{eqn_l1}) and (\ref{eqn_h}),
\begin{equation}
T = a u^i \varpi_i + \tilde{b}(\varpi^i)
\label{eqn_m}
\end{equation}
and from (\ref{eqn_b}),
\begin{equation}
{\mathcal D}_t T = D_i (-u^i T) + u^j{}_{,j}(\tilde{b}- \varpi^i \tilde{b}_{\varpi^i}) + u^i{}_{,j}\varpi^j ( a u_i +\tilde{b}_{\varpi^i} ) - \rho_{,i} a f' \varpi^i.
\label{eqn_i4}
\end{equation}
Through (\ref{id1}), we note $\rho_{,i} a f' \varpi^i = D_i (f\varpi^i)$ and $u^i{}_{,j}\varpi^j u_i = D_j (\frac{1}{2} u^i u_i \varpi^j)$ while
$u^i{}_{,j} \varpi^j \tilde{b}_{\varpi^i} = (u^{ij}+\omega^{ij})\varpi_j \tilde{b}_{\varpi^i} = u^{ij} \varpi_j \tilde{b}_{\varpi^i}$ by (\ref{id2}). 
Thus we have
\begin{equation}
{\mathcal D}_t T = 
D_i (a(\tfrac{1}{2}u^j u_j  -f)\varpi^i-u^i T ) + u^j{}_{,j}(\tilde{b}- \varpi^i \tilde{b}_{\varpi^i}) + u^{ij}\varpi_j \tilde{b}_{\varpi^i}.
\label{eqn_n}
\end{equation}

The second determining equation $0= E_{u^i} ({\mathcal D}_t T)$ thereby yields the following terms
\begin{equation}
0=A_j \varpi^j{}_{,}{}^i + B^{ij}{}_k \varpi^k{}_{,j} + mW^{kji} (\tr{u}_{,j} A_k+u^{gh}{}_{,j} B_{ghk} +\varpi^l{}_{,j}(\tr{u} A_{k\varpi^l} + u^{gh} B_{ghk\varpi^l}))
\label{eqn_aa}
\end{equation}
where
\begin{equation}
A_i = - \varpi^j \tilde{b}_{ji}, \quad
B_{ijk} = \delta_{k (i} \tilde{b}_{j)} + \varpi_{(i }\tilde{b}_{j) k}, 
\label{q}
\end{equation}
with the notation $\tilde{b}_i = \tilde{b}_{\varpi^i}$, $\tilde{b}_{ij} = \tilde{b}_{\varpi^i \varpi^j}$.
Using the identities
\begin{equation*}
\varpi^k{}_{,j} = W^k{}_{pq} \omega^{pq}{}_{,j}, \quad
u^{kl}{}_{,}{}^j =u^{jkl}+\tfrac{2}{3}\omega^{j(k}{}_,{}^{l)}, \quad
\tr{u}_,{}^j = u^{jk}{}_{,k}-\tfrac{2}{3}\omega^{kj}{}_{,k},
\end{equation*}
and collecting like terms in (\ref{eqn_aa}), we get a linear homogeneous expression in $u^{jkl}$, $\omega^{k(l}{}_,{}^{j)}$, $\omega^{k(l}{}_,{}^{j)}u^{gh}$:
\begin{eqnarray}
0 &=&
u^{jkl}W^h{}_{ji}(B_{klh}+\delta_{kl}A_h)
-\tfrac{2}{3}\omega^{lj}{}_{,}{}^k W^h{}_{ji}(B_{klh}+\delta_{kl}A_h)
+\omega^{kl}{}_{,}{}^j W^h{}_{kl}(B_{jih}+\delta_{ji}A_h) 
\nonumber \\ 
&& 
+ m u^{kl}\omega^{pq}{}_{,}{}^j W^g{}_{pq}W^h{}_{ji}(B_{klh\varpi^g}+\delta_{kl}A_{h\varpi^g}).
\label{eqn_bb}
\end{eqnarray} 
Since $\omega^{k(l}{}_,{}^{j)}$ is linearly independent of $u^{jkl}$ in $n>1$ dimensions, their coefficients in (\ref{eqn_bb}) must vanish.
From the coefficient of $\omega^{k(l}{}_,{}^{j)}u^{gh}$, we have
\begin{equation}
W^k{}_{pq}W^l{}_{ji}(B_{ghl\varpi^k}+\delta_{gh}A_{l\varpi^k}) = 0
\label{}
\end{equation} 
which yields 
\begin{equation}
B_{kli\varpi^j}=0, \quad A_{i\varpi^j} = 0. \label{eqn_cc}
\end{equation}
These two equations imply $B_{kji}$ must be a constant tensor and $A_i$ must be a constant vector.
Next, the coefficients of $u^{jkl}$ and $\omega^{k(l}{}_,{}^{j)}$ yield the equations
\begin{eqnarray}
&&W^h{}_{i(j}B_{kl)h}+W^h{}_{i(j}\delta_{kl)}A_h=0, \label{eqn_ff} \\
&&W^k{}_{gh}(B_{ijk}+\delta_{ij}A_k)-2W^k{}_{h(j}(B_{i)gk}+\delta_{i)g}A_k)=0. \label{eqn_dd}
\end{eqnarray}
We note that antisymmetrizing (\ref{eqn_ff}) in $[ij]$ leads to (\ref{eqn_dd}) after indices are renamed.
Hence only (\ref{eqn_ff}) needs to be considered.
Taking the trace of (\ref{eqn_ff}) in $(ij)$ and using the identity (\ref{id3}), we obtain
\begin{equation}
W^{hj}{}_{(k}B_{l)jh}=0. \label{eqn_gg}
\end{equation}
Such an algebraic equation can hold only as a consequence of the skew-symmetry property (\ref{prop1}), 
so (\ref{eqn_gg}) is satisfied  only if $B_{ijh} = B_{(ij)h}$ is of the form
\begin{equation}
B_{ijh}=\delta_{h(i}c_{j)} + \delta_{ij}\tilde{c}_h+\hat c_{ijh} \label{eqn_hh}
\end{equation}
where $c_i$, $\tilde{c}_h$ are constant vectors and $\hat c_{ijh} = \hat c_{(ijh)}$ is a trace-free totally symmetric constant tensor.
Hence (\ref{eqn_ff}) reduces to 
\begin{equation}
W^h{}_{i(j}\delta_{kl)}(\tilde{c}_h+A_h)+W^h{}_{i(j}\hat{c}_{kl)h}=0. \label{eqn_ii}
\end{equation}
The trace of (\ref{eqn_ii}) in $(kl)$ leads to $\tilde{c}_h+A_h=0$ (since $\hat {c}_{hk}{}^k =0$ and $W^{hk}{}_i \hat{c}_{jkh} = 0$) which implies
\begin{equation}
W^h{}_{i(j}\hat{c}_{kl)h}=0. \label{eqn_jj}
\end{equation}
By the same argument that led to (\ref{eqn_hh}), 
the only totally symmetric tensor that can satisfy (\ref{eqn_jj}) is $\hat{c}_{klh} = \delta_{(kl}\hat{c}_{h)}$, which is trace-free only if $\hat{c}_h=0$. Thus, we have
\begin{equation} 
B_{ijh}+ \delta_{ij}A_h=\delta_{h(i}c_{j)}, \quad c_j = const. \label{eqn_kk}
\end{equation}
giving the solution of equation (\ref{eqn_bb}).
Substitution of expressions (\ref{q}) into (\ref{eqn_kk}) then gives us
\begin{equation}
\delta_{k(i}\tilde{b}_{j)} + \varpi_{(i}\tilde{b}_{j)k}=\delta_{k(i}c_{j)}+\delta_{ij}\varpi^l\tilde{b}_{lk}.
\label{eqn_aaa}
\end{equation}
Antisymmetrizing (\ref{eqn_aaa}) in $[jk]$ yields
\begin{equation}
\delta_{i[k}\hat{b}_{j]}=\varpi_{[j}\tilde{b}_{k]i} \label{eqn_bbb}
\end{equation}
where $\hat{b}_j = c_j-\tilde{b}_j-2\varpi^k\tilde{b}_{kj}$.
Taking the product of (\ref{eqn_bbb}) with $\varpi_l$ antisymmetrized in $[jkl]$, followed by taking the trace in $(ik)$, we get $\hat{b}_{[j}\varpi_{l]}=0$.
This implies
\begin{equation}
\hat{b}_j = d(\varpi^i)\varpi_j \label{eqn_ccc}
\end{equation}
whence (\ref{eqn_bbb}) yields
\begin{equation}
\tilde{b}_{ki} = d(\varpi^j) \delta_{ki} + \tilde{d}(\varpi^j)\varpi_k \varpi_i \label{eqn_ddd}
\end{equation}
and thus
\begin{equation}
c_j-\tilde{b}_j = (3d+2\tilde{d}\varpi^k \varpi_k)\varpi_j. \label{eqn_dddd}
\end{equation}
Substituting (\ref{eqn_ddd}) and (\ref{eqn_dddd}) into (\ref{eqn_aaa}), we get
\begin{equation}
\tilde{d}\varpi_i \varpi_j \varpi_k = 3  (d+\tilde{d}\varpi^l \varpi_l)\delta_{(ij} \varpi_{k)} \label{eqn_eee}
\end{equation}
which directly implies
\begin{equation}
d=\tilde{d} = 0. \label{eqn_fff}
\end{equation}
From (\ref{eqn_ddd}) we thus have $\tilde{b}_{\varpi^k \varpi^i}=\tilde{b}_{ki} = 0$ and so, by direct integration,
\begin{equation}
\tilde{b} = \tilde{b}_k \varpi^k, \quad \tilde{b}_k = const.\label{eqn_hhh}
\end{equation}
where we have dropped a constant term since it does not involve $\varpi^k$.

As a result, from (\ref{eqn_m}) we have
\begin{equation}
T = a u^i\varpi_i + \tilde{b}_i \varpi^i \label{eqn_iii}
\end{equation}
with arbitrary constants $a$, $\tilde{b}_i$, while from (\ref{eqn_n}) 
\begin{equation}
{\mathcal D}T= -D_i(a(f-\tfrac{1}{2}u^j u_j)+u^i T-\tilde{b}_j u^j \varpi^i) \label{eqn_jjj}
\end{equation}
where the last term comes from
\begin{equation}
\tilde{b}_j (u^j \varpi ^i)_{,i} = \tilde{b}_j \varpi ^i u^j{}_{,i} = \tilde{b}_j \varpi _i(u^{ji}+ \omega^{ji}) = \tilde{b}_j \varpi _i u^{ij} \label{eqn_kkk}
\end{equation}
due to (\ref{id1}) and (\ref{id2}).

To conclude the proof of Theorem 3.1 and Proposition 3.2, we note that in (\ref{eqn_iii}) the term
\begin{equation}
\tilde{b}_i\varpi^i= \tilde{b}_i(W^{ijk} u_k){}_{,j}= D_j(\tilde{b}_iW^{ijk} u_k) \label{eqn_lll}
\end{equation}
is a trivial conserved density, while its corresponding moving-flux from (\ref{eqn_jjj}) is given by 
$\xi^i= -\tilde{b}_j u^j \varpi^i$, 
which fails to be divergence-free,
\begin{equation}
D_i\xi^i= -\tilde{b}_j(u^j \varpi^i){}_{,i} = -\tilde{b}_j\varpi^iu^j{}_{,i} = -\tilde{b}_j\varpi_i u^{ij} \not\equiv 0. \label{eqn_ooo}
\end{equation}

\section{Correspondence between conserved densities and Hamiltonian symmetries} \label{corresp}

The $n$-dimensional compressible Euler equations (\ref{veleqn})--(\ref{eos}) have the well-known Hamiltonian formulation \cite{Ver}
\begin{equation}
\p_t \begin{array}({c})\mathbf{u}\\\rho \end{array}=\mathcal{H}\begin{array}({c})\delta E/\delta \mathbf{u}\\ \delta E/\delta \rho \end{array},\quad 
\mathcal{H}=\begin{array}({cc})\rho^{-1}(\nabla\wedge\mathbf{u})\cdot&-\nabla\\-\nabla\cdot&0 \end{array},\quad 
E=\tfrac{1}{2}\rho|\mathbf{u}|^2 +\rho e ,
\label{hameqn}
\end{equation} 
where $e = \int\rho^{-2}P(\rho)d\rho$ is the internal energy density and $\mathcal{H}$ is called a Hamiltonian operator.
This means $\mathcal{H}$ determines a Poisson bracket \cite{Olv}
\begin{eqnarray}
\{{\mathcal F},{\mathcal G}\}_\mathcal{H} &=&
\displaystyle \int \begin{array}({cc})\delta F/\delta \mathbf{u}&\delta F/\delta \rho \end{array}\mathcal{H}\begin{array}({c})\delta G/\delta \mathbf{u}\\ \delta G/\delta \rho \end{array} d^n x 
\\
&=& \displaystyle \int \rho^{-1}(\nabla\wedge\mathbf{u})\cdot (\delta F/\delta \mathbf{u}\wedge \delta G/\delta \mathbf{u})
+\delta G/\delta \mathbf{u}\cdot\nabla \delta F/\delta \rho
-\delta F/\delta \mathbf{u}\cdot\nabla \delta G/\delta \rho\; d^n x
\nonumber
\end{eqnarray} 
having the properties that (modulo divergence terms) it is antisymmetric and obeys the Jacobi identity, for arbitrary functionals 
${\mathcal F}=\int F d^n x$ and ${\mathcal G}=\int G d^n x$ where $F$ and $G$ are functions of $t,\mathbf{x},\rho,\mathbf{u}$ and their $\mathbf{x}$-derivatives. Here $\delta /\delta \mathbf{u}$ and $\delta /\delta \rho$ denote variational derivatives, which coincide with the spatial Euler operators $E_\mathbf{u}$ and $E_\rho$ when acting on functions that do not contain time derivatives of $\rho$ and $\mathbf{u}$.

To check the Hamiltonian structure (\ref{hameqn}) produces (\ref {veleqn})--(\ref{eos}), we note $\delta E/\delta \mathbf{u}=\rho\mathbf{u}$ and $\delta E/\delta \rho=\frac{1}{2}|\mathbf{u}|^2 +e + \rho^{-1}P(\rho)$, which yields
\begin{eqnarray*}
&\rho^{-1}(\nabla\wedge\mathbf{u})\cdot \delta E/\delta \mathbf{u}=(\nabla\wedge\mathbf{u})\cdot\mathbf{u}=|\mathbf{u}|\nabla|\mathbf{u}|-\mathbf{u}\cdot\nabla\mathbf{u}, & 
\\
&-\nabla\cdot\delta E/\delta \mathbf{u}=-\nabla\cdot(\rho\mathbf{u}),& 
\\
&-\nabla\delta E/\delta \rho=-\tfrac{1}{2}\nabla|\mathbf{u}|^2-\nabla(e + \rho^{-1}P(\rho))=-|\mathbf{u}|\nabla|\mathbf{u}|-\rho^{-1}P'(\rho)\nabla\rho,& 
\end{eqnarray*}
whence we obtain 
\begin{eqnarray}
\mathcal{H}\begin{array}({c})\delta E/\delta \mathbf{u}\\ \delta E/\delta \rho \end{array}=&\begin{array}({c})\rho^{-1}(\nabla\wedge\mathbf{u})\cdot\delta E/\delta \mathbf{u}-\nabla\delta E/\delta \rho\\-\nabla\cdot\delta E/\delta \mathbf{u} \end{array} \nonumber \\
=&\begin{array}({c})-\mathbf{u}\cdot\nabla\mathbf{u}-\rho^{-1}\nabla P(\rho)\\-\nabla\cdot(\rho\mathbf{u})\end{array}=\begin{array}({c})\p_t \mathbf{u}\\\p_t \rho \end{array}.
\label{hameqn1}
\end{eqnarray} 

Through this formulation the Hamiltonian operator $\mathcal{H}$ gives rise to an explicit mapping that produces symmetries of the compressible Euler equations from conservation laws as follows.
Recall, fluid symmetries \cite{Ibr} are described by an infinitesimal transformation
\begin{equation}
\hat{X}\mathbf{u}=\hat{\boldsymbol\eta},\quad \hat{X}\rho=\hat\eta,
\label{symm}
\end{equation} 
on all formal solutions of (\ref {veleqn})--(\ref{eos}), where $\hat{\boldsymbol\eta}$ and $\hat\eta$ are some functions of $t,\mathbf{x},\rho,\mathbf{u}$, and $\mathbf{x}$-derivatives of $\rho$, $\mathbf{u}$ determined by infinitesimal invariance \cite{Olv,BA} of the Euler equations (\ref {veleqn})--(\ref{eos}): 
\begin{equation*}
D_t\hat{\boldsymbol\eta}+\mathbf{u}\cdot\nabla\hat{\boldsymbol\eta}+\hat{\boldsymbol\eta}\cdot\nabla\mathbf{u}+\nabla(\rho^{-1}P'(\rho)\hat\eta)=0, \quad
D_t\hat\eta+\nabla\cdot(\hat\eta\mathbf{u}+\rho\hat{\boldsymbol\eta})=0.
\end{equation*}
Now if $T$ is a conserved density of the Euler equations (\ref {veleqn})--(\ref{eos}) then the mapping
\begin{equation}
-\mathcal{H}\begin{array}({c})\delta T/\delta \mathbf{u}\\
\delta T/\delta \rho\end{array}=\hat{X}\begin{array}({c})\mathbf{u}\\\rho \end{array}
\label{shameqn}
\end{equation} 
can be shown (cf. general results in \cite{Olv}) to yield a symmetry 
$\hat{X}= \hat{\boldsymbol\eta}\rfloor\p_\mathbf{u}+\hat{\eta}\p_{\rho}$, 
given by 
\begin{equation}
\hat{\boldsymbol\eta} =-\rho^{-1}(\nabla\wedge\mathbf{u})\cdot \delta T/\delta \mathbf{u}+ \nabla\delta T/\delta \rho, \quad
\hat{\eta} =\nabla\cdot\delta T/\delta \mathbf{u}.
\end{equation}
In particular, as seen from (\ref {hameqn1}), the conserved energy density (\ref{kin_e}) yields a time translation symmetry
\begin{eqnarray}
T=\tfrac{1}{2}\rho|\mathbf{u}|^2+\rho e \to \hat{X}\mathbf{u}=-\p_t\mathbf{u}, \quad \hat{X}\rho=-\p_t\rho.
\label{Ta}
\end{eqnarray}

For the other kinematic conserved densities listed in Theorem 2.1, we find the following correspondences:
mass density (\ref{kin_a}) yields a trivial symmetry
\begin{eqnarray}
T=\rho \to \hat{X}\mathbf{u}=0=\hat{X}\rho;
\label{Tb}
\end{eqnarray}
momentum densities (\ref{kin_b}) yield space translation symmetries
\begin{eqnarray}
T=\rho\mathbf{u} \to \hat{X}\mathbf{u}=\nabla\mathbf{u},\quad \hat{X}\rho=\nabla\rho;
\label{Tc}
\end{eqnarray}
angular momentum densities (\ref{kin_c}) yield rotation symmetries
\begin{eqnarray}
T=\rho\mathbf{u}\wedge\mathbf{x} \to \hat{X}\mathbf{u}=(\mathbf{x}\wedge\nabla)\mathbf{u}-\mathbf{u}\wedge\mathbf{g},\quad \hat{X}\rho=\mathbf{x}\wedge\nabla\rho;
\label{Td}
\end{eqnarray}
Galilean momentum densities (\ref{kin_d}) give rise to Galilean boost symmetries
\begin{eqnarray}
T=\rho(t\mathbf{u}-\mathbf{x})\to \hat{X}\mathbf{u}=t\nabla\mathbf{u}-\mathbf{g}, \quad \hat{X}\rho=t\nabla\rho.
\label{Te}
\end{eqnarray}
Here $\mathbf{g}$ is Euclidean metric tensor on $\mathbb R^n$ 
(recall, in Cartesian coordinates, the components of $\mathbf{g}$ are given by the Kronecker symbol).

In the case of a polytropic \eos/ (\ref{polytropic}) with special exponent $\gamma=1+2/n$, the similarity energy (\ref{kin_f}) yields
\begin{eqnarray}
&T=tE-\tfrac{1}{2}\rho(\mathbf{u}\cdot\mathbf{x}) \to &\hat{X}\mathbf{u}=-(t\p_t\mathbf{u}+\tfrac{1}{2}(\mathbf{x}\cdot\nabla)\mathbf{u}+\tfrac{1}{2}\mathbf{u}), 
\nonumber\\ && 
\hat{X}\rho=-(t\p_t\rho+\tfrac{1}{2}(\mathbf{x}\cdot\nabla)\rho+\tfrac{1}{2}n\rho)
\label{Tf}
\end{eqnarray}
which is a scaling (similarity) symmetry;
and the dilational energy (\ref{kin_g}) yields
\begin{eqnarray}
&T=t^2E-t\rho(\mathbf{u}\cdot\mathbf{x})+\tfrac{1}{2}\rho|\mathbf{x}|^2
\to &\hat{X}\mathbf{u}= -(t^2\p_t\mathbf{u}+t(\mathbf{x}\cdot\nabla)\mathbf{u}+t\mathbf{u}-\mathbf{x}), \nonumber \\
&&\hat{X}\rho = -(t^2\p_t\rho+t(\mathbf{x}\cdot\nabla)\rho+nt\rho)
\label{Tg}
\end{eqnarray}
which we call a Galilean dilation symmetry because it preserves $\mathbf{x}-t\mathbf{u}$ and $\mathbf{x}/t$. 
Here $E$ is polytropic energy density (\ref{polyenergy}).

In contrast, for the helicity and generalized enstrophy densities (\ref{vconlaw1})--(\ref{vconlaw2}), we get:
\begin{eqnarray}
T=\mathbf{u}\cdot{\boldsymbol\varpi}\to \hat{X}\mathbf{u}=0=\hat{X}\rho; \label{Th}\\
T=\rho f(\varpi/\rho)\to \hat{X}\mathbf{u}=0=\hat{X}\rho. \label{Ti}
\end{eqnarray}
Thus we have the following classification result.

\textbf{Proposition 4.1:} 
In all dimensions $n>1$, the nontrivial infinitesimal symmetries produced from the kinematic conserved densities (\ref{kin_a})--(\ref{kin_g}) under the Hamiltonian mapping (\ref{shameqn}) consist of space translations (\ref{Tc}), rotations (\ref{Td}), Galilean boosts (\ref{Te}), and a time translation (\ref{Ta}) for general \esos/, plus a similarity scaling (\ref{Tf}) and a  Galilean dilation (\ref{Tg}) for the special polytropic \eos/ (\ref{special}). 

All of these Hamiltonian symmetries (\ref{Ta})--(\ref{Ti}) can be seen to have the form of infinitesimal point transformations
\begin{equation}
\hat{\boldsymbol\eta}={\boldsymbol\eta}-\tau \p_t\mathbf{u}-{\boldsymbol\xi}\cdot\nabla \mathbf{u}, \quad
\hat\eta=\eta-\tau \p_t\rho -{\boldsymbol\xi}\cdot\nabla\rho
\label{Hsymm}
\end{equation}
so thus
\begin{equation}
\hat{X}= \hat{\boldsymbol\eta}\rfloor\p_\mathbf{u}+\hat{\eta}\p_{\rho} \leftrightarrow X= \tau \p_t +{\boldsymbol\xi}\rfloor\p_\mathbf{x}+{\boldsymbol\eta}\rfloor\p_\mathbf{u}+\eta\p_\rho
\end{equation}
where ${\boldsymbol\eta}$, $\eta$ are functions of $t$, $\mathbf{x}$, $\mathbf{u}$, $\rho$, while $\tau$, ${\boldsymbol\xi}$ are functions only of $t$, $\mathbf{x}$. 
These symmetries have the following geometrical description (proven in section 5).

\textbf{Proposition 4.2:} 
Let ${\boldsymbol\zeta}(\mathbf{x})$ be any solution of the dilational Killing vector equation
${\mathcal L}_{\boldsymbol\zeta}\mathbf{g} = \Omega \mathbf{g}$, $\Omega = const.$, on $\mathbb R^n$, and let ${\boldsymbol\chi}(\mathbf{x})$ be any irrotational solution of the same equation, 
${\mathcal L}_{\boldsymbol\chi}\mathbf{g} = \tilde{\Omega} \mathbf{g}$, $\tilde\Omega = const.$, $\nabla \wedge {\boldsymbol\chi} = 0$, 
on $\mathbb R^n$.
Then the Hamiltonian symmetries (\ref{Tb})--(\ref{Tg}) corresponding to the kinematic conserved densities (\ref{kin_b}),(\ref{kin_c}),(\ref{kin_d}),(\ref{kin_f}),(\ref{kin_g}) 
have the form
\begin{eqnarray}
&X=  {\boldsymbol\zeta} \rfloor \p_\mathbf{x} + \Omega t\p_t + \tfrac{1}{2}n \Omega \rho \p_\rho +(\Omega \mathbf{u} + \tfrac{1}{2}(\nabla\wedge{\boldsymbol\zeta})\cdot\mathbf{u})\rfloor \p_\mathbf{u},& \\
&X=  t{\boldsymbol\chi} \rfloor \p_\mathbf{x}  + \tfrac{1}{2}\tilde\Omega t^2\p_t +\tfrac{1}{2}n \tilde\Omega t\rho \p_\rho +(\tfrac{1}{2}t\tilde\Omega \mathbf{u} -{\boldsymbol\chi})\rfloor \p_\mathbf{u}.& 
\end{eqnarray} 

A comparison of these Hamiltonian symmetries with all of the well-known point symmetries \cite{Ovs,Ibr} admitted by the compressible Euler equations (\ref {veleqn})--(\ref{eos}) in dimensions $n>1$ gives the following classification results.

For general \esos/ (\ref{eos}):\\
\noindent \begin{tabular}{|c|c|c|c|c|}
\hline
Point Symmetry $X$   &  Description & Number  & Hamiltonian \\
&&& Correspondence\\
\hline
$\p_t$ & Time translation & $1$ & Energy\\
$\p_{\mathbf{x}}$ & Space translations & $n$ & Momentum\\
$\mathbf{x}\wedge\p_{\mathbf{x}}+\mathbf{u}\wedge\p_{\mathbf{u}}$ & Rotations & $n(n-1)/2$ & Angular momentum\\
$t\p_{\mathbf{x}}+\p_{\mathbf{u}}$ & Galilean boost & $n$ & Galilean momentum\\
$t\p_{t}+\mathbf{x}\rfloor\p_{\mathbf{x}}$ & Dilation & 1 & Nil\\
\hline
\end{tabular}\\

For polytropic \esos/ (\ref{polytropic}):\\
\noindent \begin{tabular}{|c|c|c|c|c}
\hline
Point Symmetry $X$   &  Description  & Exponent &Hamiltonian \\
&&& Correspondence\\
\hline
$\mathbf{x}\rfloor\p_{\mathbf{x}}+\mathbf{u}\rfloor\p_{\mathbf{u}}+\frac{2}{\gamma-1}\rho\p_{\rho}$ & Scaling & $\gamma \neq1$ & Nil\\
$t\p_t+\frac{1}{2}\mathbf{x}\rfloor\p_\mathbf{x}-\frac{1}{2}\mathbf{u}\rfloor\p_\mathbf{u}-\frac{n}{2}\rho \p_\rho$& Similarity Scaling & $\gamma=1+2/n$ & Similarity energy\\
$t^2\p_t+t\mathbf{x}\rfloor\p_\mathbf{x}+(\mathbf{x}-t\mathbf{u})\rfloor\p_\mathbf{u}-nt\rho\p_\rho$ & Galilean dilation & " & Dilational energy\\
	\hline
\end{tabular}\\

Finally, we remark that conserved densities $T$ with the property
\begin{eqnarray}
\mathcal{H}\begin{array}({c})\delta T/\delta \mathbf{u}\\ \delta T/\delta \rho \end{array}=0
\end{eqnarray}
are known as a {\it Hamiltonian Casimir}. 
Such conserved densities are distinguished by having no correspondence to any symmetry of the Euler equations (\ref {veleqn})--(\ref{eos}).
Our classification shows that the only vorticity Casimirs and kinematic Casimirs admitted by the fluid Hamiltonian (\ref{hameqn}) consist of the conserved densities for helicity (\ref{vconlaw1}), enstrophy (\ref{vconlaw2}), and mass (\ref{kin_a}).

\section{Summary and Concluding Remarks}

For the Euler equations (\ref {veleqn})--(\ref{eos}) governing compressible isentropic fluid flow in $n>1$ dimensions, we have directly classified all nontrivial kinematic and vorticity conservation laws (\ref{conlaw}) by solving the determining equations (\ref{deteqn}) for conserved densities of the respective forms (\ref {mech}) and (\ref{vorticity}).
Alternatively, our classification of conservation laws can be carried out by means of multipliers \cite{AB1997,AB2002a,AB2002b,BCA}.
This approach is most easily presented as follows using index notation (cf. section 2.2).

Let $T(t,x^j,\rho,u^k,\rho_{,j},u^k{}_{,j},\ldots)$ 
be a nontrivial conserved density and let 
$X^i(t,x^j,\rho,u^k,$
$\rho_{,j},u^k{}_{,j},\ldots)$ 
be a spatial flux vector, 
given by some functions of the time and space coordinates $t$ and $x^i$, the fluid density $\rho$ and fluid velocity $u^i$, and their spatial derivatives $\rho_{,i}$, $u^j{}_{,i}$, etc. with respect to $x^i$, 
which satisfy a local continuity equation
\begin{equation}
{\mathcal D}_t T + D_i X^i = 0
\label{cr_a}
\end{equation}
where
\begin{equation}
{\mathcal D}_t = \p_t - (\rho u^i)_{,i} \p_\rho - (u^j u^i{}_{,j} + \rho^{-1} P'(\rho)\rho_,{}^i)\p_{u^i} + \cdots
\label{cr_a1}
\end{equation}
is the time derivative defined through the Euler equations (\ref {eqn_11}).
If we express (\ref{cr_a}) in terms of the total derivative $D_t=\p_t+\rho_t \p_\rho + u^i_t \p_{u^i}+\cdots$, 
then we obtain an equivalent equation
\begin{equation}
D_t T + D_i \tilde{X}^i = (\rho_t + (\rho u^i)_{,i})Q +(u^i_t+u^j u^i{}_{,j} + \rho^{-1} P'(\rho ) \rho{}_,{}^i) Q_i
\label{cr_b}
\end{equation}
holding for $\rho$ and $u^i$ given by {\it arbitrary} functions of $t$ and $x^i$, where
\begin{equation}
Q= E_\rho(T), \quad Q_i = E_{u^i}(T)
\label{cr_c}
\end{equation}
are functions of $t$, $x^j$, $\rho$, $u^k$, $\rho_{,j}$, $u^k{}_{,j}$ etc.,
and where $\tilde{X}^i$ differs from $X^i$ by terms that are linear homogeneous in the Euler equations (\ref {eqn_11}) and total spatial derivatives of (\ref {eqn_11}).
This equation (\ref{cr_b}) is called the {\it characteristic form} of the conservation law (\ref{cr_a}).
It establishes, firstly, that every nontrivial local conservation law of the Euler equations (\ref {eqn_11}) arises from {\it multipliers} (\ref{cr_c}).
Secondly, since the spatial Euler operators $E_\rho$ and $E_{u^i}$ annihilate divergences $D_i \Theta^i$ for any vector function $\Theta^i(t$, $x^j$,$\rho$,$u^k$,$\rho_{,j}$,$u^k{}_,{}_j,\ldots)$, the relation (\ref{cr_c}) shows that any two local conservation laws differing by a trivial conserved density of the form $D_i \Theta^i$ have the same multipliers.
Thus, there is a one-to-one correspondence between nontrivial conserved densities (modulo spatial divergences) and non-zero multipliers.

Necessary and sufficient equations for determining multipliers \cite{AB2002b,BCA} are given by applying variational derivative operators $\delta/\delta \rho$ and $\delta/\delta u^i$ to the characteristic equation (\ref{cr_b}),
yielding a linear homogeneous polynomial system in $\rho_t$, $u^i_t$, $\rho_{t,j}$, $u^i_{t,j}$, etc. whose coefficients must separately vanish. 
The resulting determining equations for $Q$($t$,$x^j$,$\rho$,$u^k$,$\rho_{,j}$,$u^k{}_,{}_j$,$\ldots)$ and $Q_i$($t$, $x^j$,$\rho$,$u^k$,$\rho_{,j}$,$u^k{}_,{}_j$,$\ldots)$ consist of
\begin{eqnarray}
&&{\mathcal D}_t Q + u^i D_i Q + f'(\rho) D^i Q_i = 0, \label{cr_d1}\\
&&{\mathcal D}_t Q_i + u^j D_j Q_i + 2 u^j{}_{,[j} Q_{i]} + \rho D^i Q = 0, \label{cr_d2}\\
&&{\mathcal L}_{u^j}Q_i = {\mathcal L}^*_{u^i} Q_j , \quad {\mathcal L}_\rho Q = {\mathcal L}^*_\rho Q, \quad
{\mathcal L}_{u^i}Q = {\mathcal L}^*_\rho Q_i , \quad {\mathcal L}_\rho Q_i = {\mathcal L}^*_{u^i} Q ,\label{cr_d4}
\end{eqnarray}
where ${\mathcal L}_{u^i}$ and ${\mathcal L}_\rho$ denote linearization operators (Frechet derivatives) with respect to $u^i$ and $\rho$; ${\mathcal L}^*_{u^i}$ and ${\mathcal L}^*_\rho$ denote the adjoint linearization operators \cite{AB2002b,BCA}.
We now note, first, (\ref{cr_d4}) provides the necessary and sufficient conditions \cite{Olv,BCA} for $Q$ and $Q_i$ to have the form of variational derivatives of some function with respect to $\rho$ and $u^i$.
(Moreover, this function can be constructed explicitly from $Q$ and $Q^i$ by means of homotopy integral formulas \cite{Olv,AB2002b,BCA} 
or by an algebraic scaling formula \cite{Anc} based on invariance of the Euler equations under dilations $t\to \lambda t$, $x^i\to \lambda x^i$.)
Second, we note (\ref{cr_d1}) and (\ref{cr_d2}) constitute the adjoint of the determining equations for symmetries 
\begin{eqnarray}
&&
{\mathcal D}_t\hat\eta + D_i(u^i\hat\eta+\rho\hat\eta^i)=0, \\
&&
{\mathcal D}_t\hat\eta^i + u^j D_j\hat\eta^i + u^i{}_{,j}\hat\eta^j + D_i(f'(\rho)\hat\eta )=0,
\end{eqnarray}
where
\begin{equation}
\hat{X}=\hat\eta (t, x^j,\rho,u^k,\rho_{,j},u^k{}_,{}_j,\ldots) \p_\rho + \hat{\eta}^i(t, x^j,\rho,u^k,\rho_{,j},u^k{}_,{}_j,\ldots)\p_{u^i}
\label{cr_e}
\end{equation}
is the symmetry generator in characteristic form.
Thus, multipliers can be characterized as adjoint-symmetries that have a variational form \cite{AB2002b,BCA}.
In particular, this formulation reduces the determination of multipliers and hence of conservation laws to an adjoint version of the determination of symmetries.

We now list the multipliers for the kinematic conservation laws (\ref{kin_a})--(\ref{kin_g}) and vorticity conservation laws (\ref{vconlaw1})--(\ref{vconlaw2})
in the following two tables. 

For general \esos/ (\ref{eos}):\\
\noindent \begin{tabular}{|c|c|c|c|}
\hline
Conserved density $T$   &  Description & $Q=\delta T/\delta \rho$  &  $Q_i=\delta T/\delta u^i$  \\
\hline
$\rho$ & Mass & $1$ & $0$ \\
$\rho u^k$ & Momentum & $u^k$ & $\rho \delta^k_i$ \\
$\rho(u^jx^k - u^kx^j)$ &Angular momentum & $u^jx^k - u^kx^j$ & $\rho (x^k\delta^j_i-x^j\delta^k_i)$ \\
$\rho(tu^k-x^k)$ & Galilean momentum & $tu^k-x^k$ & $\rho t\delta^k_i$  \\
$\frac{1}{2}\rho u^ku_k+\rho \int \rho^{-2} P(\rho)d\rho$ & Energy & $\frac{1}{2}u^ku_k + \int \rho^{-1}P'(\rho)d\rho$ & $\rho u_i$ \\
$u^k \varpi_k $		& Helicity $(n=2m+1)$& $0$ & $(m+1)\varpi_i$ \\
$\rho f(\varpi/\rho) $ & Enstrophy $(n=2m)$& $f(\varpi/\rho)-\rho^{-1}\varpi f'(\varpi/\rho)$ & $m w_{ij}f''(\varpi/\rho) \rho_,{}^j$ \\
\hline  
\end{tabular}\\

For the distinguished polytropic \eos/ (\ref{special}):\\
\noindent \begin{tabular}{|c|c|c|c|}
\hline
Conserved density $T$   &  Description & $Q=\delta T/\delta \rho$  &  $Q_i=\delta T/\delta u^i$  \\
\hline
$\frac{1}{2}\rho u^ku_k+\frac{n\kappa}{2}\rho^{1+\frac{2}{n}}=E$ & Polytropic energy & $\frac{1}{2}u^ku_k +(1+ \frac{n}{2})\kappa\rho^{\frac{2}{n}}$ & $\rho u_i$ \\
$tE-\frac{1}{2}\rho u^kx_k$ & Similarity energy & $\frac{1}{2}u^k(tu_k - x_k)+ t(1+ \frac{n}{2})\kappa\rho^{\frac{2}{n}}$ & $\rho(tu_i-\frac{1}{2} x_i)$ \\
$t^2E-\rho (tu^k-\frac{1}{2} x^k)x_k$ & Dilational energy & $\frac{1}{2}(tu_k - x_k)(tu^k - x^k)+ t^2(1+ \frac{n}{2})\kappa\rho^{\frac{2}{n}}$ & $t\rho(t u_i-x_i)$ \\
\hline
\end{tabular}\\ 

The adjoint relation between multipliers and symmetries can be expressed in an explicit form through the Hamiltonian formulation of the Euler equations (\ref{hameqn}):
\begin{equation}
\begin{array}({c})\hat\eta^i\\\hat\eta \end{array}=-\mathcal{H}\begin{array}({c}) Q^j \\ Q \end{array} ,\quad
-\mathcal{H}= \begin{array}({cc})2\rho^{-1}u^{[i}{}_,{}^{j]} & D^i \\D^j&0\end{array}
\label{hamcorr}
\end{equation}
where $\mathcal{H}$ is the Hamiltonian operator. 
As shown in section 4, the mapping defined by (\ref{hamcorr}) annihilates the multipliers for Hamiltonian Casimirs consisting of the conserved densities for mass, helicity and enstrophy. 
From the specific form of the multipliers for the remaining conserved densities---momentum, angular momentum, Galilean momentum, energy, similarity energy and dilational energy ---given in the preceding two tables, we find
\begin{equation}
Q = \tau\big( \tfrac{1}{2} u^i u_i + \int \rho^{-1}P'(\rho)d\rho\big) 
- \xi_i u^i + \sigma, \quad Q_i = \rho (\tau u_i - \xi_i )
\label {hmap1}
\end{equation}
and
\begin{equation}
-\hat\eta = (\xi^i \rho){}_,{}_i +\tau \rho_t, \quad -\hat\eta^i = \xi_j u^i{}_,{}^j  + \xi_{j,}{}^i u^j  + \tau u^i_t - \sigma_,{}^i= (\xi_j u^j - \sigma ){}_,{}^i + 2\xi_j \omega^{ji} + \tau u^i_t ,
\label {hmap2}
\end{equation}
where $\xi_i$, $\tau$, $\sigma$ are shown in the following table.

\noindent \begin{tabular}{|c|c|c|c|c|c|}
\hline
 $\xi_i$ & $\tau$  & $\sigma$ & Symmetry & Conserved density & Equation of State  \\
\hline
$-\delta^j_i$ & $0$ & $0$ & Space translations& Momentum & general\\
$-2\delta_i^{[j} x^{k]}$ & $0$ & $0$  & Rotations & Angular momentum & "\\
$-t\delta^j_i$ & $0$ & $-x^j$  & Galilean boosts & Galilean momentum & "\\
$0$ & $1$ & $0$  & Time translation  & Energy& "\\
$\frac{1}{2}x_i$ & $t$ & $0$  & Similarity scaling & Similarity energy& polytropic $\gamma=1+2/n$ \\
$tx_i$ & $t^2$ & $\frac{1}{2}x^jx_j$  & Galilean dilation scaling & Dilational energy& "\\
\hline
\end{tabular}\\

It is straightforward to show that $\xi_i$, $\tau$, $\sigma$ satisfy the following system of equations:
\begin{equation}
2 \xi_{(i,j)} = \tau_t\delta_{ij}, \quad \xi_{it} = \sigma_{,i}, \quad \sigma_t = 0, \quad \tau_{,i} = 0.
\end{equation}
In particular, we note 
\begin{equation}
\xi_i = \zeta_i +t\chi_i, \quad \tau = c+ (2/n) t \zeta_{i,}{}^i +(1/n) t^2 \chi_{i,}{}^i, \quad \sigma = \int \chi_i dx^i
\label{eqn186}
\end{equation}
with $c=const.$, where
\begin{equation}
\zeta_{(i,j)} = \tfrac{1}{2} \Omega \delta_{ij}, \quad \Omega = const.
\end{equation}
is the equation defining dilational Killing vectors $\zeta_i(x^j)$ on $\mathbb{R}^n$, and 
\begin{equation}
\chi_{[i,j]} = 0, \quad \chi_{(i,j)} = \tfrac{1}{2} \tilde\Omega \delta_{ij}, \quad \tilde\Omega = const.
\label{eqn188}
\end{equation}
are the equations defining irrotational dilational Killing vectors $\chi_i(x^j)$ on $\mathbb{R}^n$. As a result, the symmetries corresponding to the multipliers for the non-Casimir conserved densities under the Hamiltonian mapping (\ref{hmap1})--(\ref{hmap2}) have the form of geometrical point symmetries 
\begin{eqnarray*}
X= \big((2/n)\int \xi^j{}_{,j} dt\big)\p_t + \xi^i \p_{x^i} + \xi^j{}_{,j} \rho \p_\rho + \big((1/n)\xi^j{}_{,j} u^i + \xi^{[j}{}_,{}^{i]} u_j - \xi^i_t \big) \p_{u^i}
\end{eqnarray*}
given in terms of $\xi^i(t,x^j)$ through (\ref{eqn186})--(\ref{eqn188}). 

\end{document}